\documentclass[universe,article,preprint
,pdftex,moreauthors]{Definitions/mdpi} 
\usepackage{isotope}
\usepackage{slashed}
\usepackage{amsmath}
\usepackage{url,color,colordvi,epsfig,pstricks}
\usepackage{graphicx,bm,dcolumn}
\usepackage{physics}
\usepackage{inputenc}
\usepackage{mathtools}

\newcommand{\Pv}{{\bf P}}

\graphicspath{Plots}

\firstpage{1} 
\pubvolume{1}
\issuenum{1}
\articlenumber{0}
\pubyear{2023}
\copyrightyear{2023}
\datereceived{ } 
\daterevised{ } 
\dateaccepted{ } 
\datepublished{ } 
\hreflink{https://doi.org/} 

\Title{Lepton-Nucleus Interactions within Microscopic Approaches}

\TitleCitation{Lepton-Nucleus Interactions within Microscopic Approaches}

\Author{Alessandro Lovato $^{2,3}$, Alexis Nikolakopoulos $^{1}$,  Noemi Rocco$^{1}$, and Noah Steinberg$^{1}$}

\AuthorNames{Alexis Nikolakopoulos, Noah Steinberg, Alessandro Lovato , Noemi Rocco}

\AuthorCitation{Nikolakopoulos, A.; Steinberg, N.; Lovato , A.; Rocco, N.}

\address{%
$^{1}$ \quad Theoretical Physics Department, Fermi National Accelerator Laboratory, P.O. Box 500, Batavia, IL 60410, USA\\
$^{2}$ \quad Physics Division, Argonne National Laboratory. Argonne, Illinois 60439, USA;
lovato@alcf.anl.gov \\
$^{3}$ \quad INFN, Trento Institute of Fundamental Physics and Applications, 38123 Trento, Italy}

\begin{document}

\abstract{
{ This review paper emphasizes the significance of microscopic calculations with quantified theoretical error estimates in studying lepton-nucleus interactions and their implications for electron-scattering and accelerator neutrino-oscillation measurements. We investigate two approaches: Green's Function Monte Carlo and the extended factorization scheme, utilizing realistic nuclear target spectral functions.\\
In our study, we include relativistic effects in Green's Function Monte Carlo and validate the inclusive electron-scattering cross section on carbon using available data. We compare the flux folded cross sections for neutrino-Carbon scattering with T2K and MINER$\nu$A experiments, noting the substantial impact of relativistic effects in reducing the theoretical curve strength when compared to MINER$\nu$A data. Additionally, we demonstrate that quantum Monte Carlo-based spectral functions accurately reproduce the quasi-elastic region in electron-scattering data and T2K flux folded cross sections.\\
By comparing results from Green's Function Monte Carlo and the spectral function approach, which share a similar initial target state description, we quantify errors associated with approximations in the factorization scheme and the relativistic treatment of kinematics in Green's Function Monte Carlo.}
}



\section{Introduction}
The electron-scattering experimental program at Jefferson laboratory aimed at investigating nuclear short-range correlations~\cite{Patsyuk:2019ocp,Fomin:2017ydn,CLAS:2022odn}, and the accelerator neutrino program, which will culminate with the completion of DUNE~\cite{DUNE:2022aul}, have been the springboard for significant progress in theoretical calculations of lepton-nucleus scattering. Approaches based on empirical effective nucleon-nucleon interactions~\cite{WAROQUIER1983269, HOROWITZ1981, SHARMA1993, RevModPhys.75.121}  have been used to study inclusive and semi-inclusive neutrino scattering data in a variety of kinematic setups~\cite{Franco-Patino:2022tvv, Amaro:2019zos, Amaro:2021sec, Gonzalez-Jimenez:2019qhq, Pandey:2016, Bourguille:2020bvw, PhysRevC.68.048501, Dolan:2019bxf, Tom:SRC, Martini:Ar, Jachowicz:2019JPG, PhysRevC.75.034613}. Despite their success, it is still imperative to attain a description of lepton-nucleus scattering from microscopic nuclear dynamics, which assumes that the structure and electroweak properties of atomic nuclei can be modeled in terms of nuclear potentials and consistent electroweak currents. These microscopic approaches allow one to quantify the theoretical uncertainties due to both modeling nuclear dynamics and solving the many-body Schr\"odinger equation. This aspect is critical for a meaningful comparison with electron-scattering data, and, perhaps more importantly, to rigorously assess the error budget of neutrino-oscillation parameters. Moreover, retaining nuclear correlations in the initial target state is important to explain the observed abundances of neutron-proton correlated pairs with respect to the proton-proton and neutron-neutron ones~\cite{CLAS:2020rue}. These experimental measurements can in turn shed light on the behavior of nuclear forces at short distances, which plays an important role in the equation of state of infinite nuclear matter at high density~\cite{Benhar:2021doc,Lovato:2022apd}. 

Variational Monte Carlo (VMC) and Green's Function Monte Carlo (GFMC) methods have proven to be extremely successful for computing the structure and electroweak transitions of atomic nuclei taking as input highly-realistic nuclear Hamiltonians~\cite{Carlson:2014vla}. Over the past decade, these methods have been employed to carry out microscopic calculations of the electroweak response functions of light nuclei, fully retaining correlations and consistent one- and two-body currents~\cite{Lovato:2013cua,Lovato:2014eva,Lovato:2016gkq,Lovato:2020kba}. Computing the hadronic response tensor is a highly-nontrivial task, as it involves transitions to the initial ground-state of the target to excited states, both bound and in the continuum. The prohibitive difficulties involved in computing all transitions mediated by the electroweak current operators are circumvented by employing integral-transform techniques. Within this approach, the electroweak response functions are inferred from their Laplace transforms, denoted as Euclidean responses, that are estimated during the GFMC imaginary time propagation. Retrieving the energy dependence of the response functions from their Euclidean counterparts is nontrivial. The maximum entropy method~\cite{Bryan:1990,Jarrell:1996rrw} has
been extensively employed to retrieve the energy dependence of the electroweak response functions in the smooth quasi-elastic region. More recently, inversion methods based on deep-neural networks have been proposed as viable alternatives and seem to be more accurate especially in the low-energy transfer region~\cite{Raghavan:2020bze}.

One of the main limitations of the GFMC approach lies in the nonrelativistic formulation of the many-body problem. Although the leading relativistic corrections are included in the transition operators~\cite{Shen:2012xz}, the kinematics of the reaction is nonrelativistic, thereby limiting the application of the GFMC to moderate values of the momentum transfer. This restriction is particularly relevant when making predictions for inclusive neutrino-nucleus cross sections since the incoming neutrino flux is not monochromatic and its tails extend to high energies. In Refs.~\cite{Rocco:2018,Nikolakopoulos:2023zse} relativistic effects in GFMC calculations of lepton-nucleus scattering are controlled by choosing a reference frame which minimizes nucleon momenta and utilizing the so-called ``two-fragment'' model to include relativity in the kinematics of the reaction. 

On the other hand, alternative approaches based on the factorization of the nuclear final state, such as the spectral function (SF) formalism~\cite{Rocco:2018mwt}, can reach larger energies and momentum transfers, as they include relativistic effects in both the kinematics and in the interaction vertex. In contrast with the GFMC, the SF approach can access exclusive channels and larger nuclei. However, while based on a similar treatment of the initial target state, factorizing the final state involves additional approximations, which are only valid at large momentum transfer, whose validity can be tested against comparisons with GFMC calculations~\cite{Rocco:2016ejr,Simons:2022ltq}. 

In this work, we first review the GFMC and SF approaches to compute inclusive lepton-nucleus scattering, placing particular emphasis on the role of relativistic effects and two-body currents. We then compare the SF predictions for the neutrino-nucleus cross-sections and compare with the MINER$\nu$A Medium Energy charge-current quasielastic (CCQE)-like data~\cite{MINERvA:2023kuz}. Finally, we
present unpublished GFMC calculations for the inclusive electron-$^{12}$C cross sections that include relativistic corrections. 

\section{Methodology}

The lepton-nucleus differential cross section in the one-boson exchange approximation can be written as
\begin{equation}
    \left(\frac{d\sigma}{dE'd\Omega'}\right)_{l} = \mathcal{C}_{l}L^{l}_{\mu\nu}R^{\mu\nu}\, ,
    \label{eq:cross_sec}
\end{equation}
where $l$ stands for either a charged lepton or neutrino, $C_{l}$ is a coupling term, and $E^\prime$ and $\Omega^\prime$ are the energy and solid angle of the lepton in the final state. The leptonic tensor is denoted by $L^{l}_{\mu\nu}$ and is a function of the initial and final lepton four-momenta $k$ and $k^\prime$, respectively. {
 For small lepton energy the Coulomb distortion of the outgoing lepton in the potential of the residual nucleus can be described multiplying the cross section by the Fermi function $F(Z,k^\prime)$ with $Z$ denoting the number of protons. The expression of this function for charge-raising reactions is given in Ref.~\cite{Shen:2012xz}  and it is equal to one otherwise.
For higher energies, the correction is provided by the modified effective momentum approximation as discussed in Ref.~\cite{Engel:1997fy}, where an effective momentum is utilized for the final lepton 
correcting its value with the Coulomb energy evaluated at the center of the nucleus, and modifying the phase space representing the density of final states accordingly. For the comparisons with T2K and MINER$\nu$A data discussed in this review, the effect of Coulomb corrections is negligible as discussed in Fig.9 of Ref.~\cite{Jachowicz:2021ieb} and therefore they have not been included.}

In this review, we will consider electromagnetic and CC electroweak interactions.
In the first case, we have an electron in both the initial and final state, the prefactor reads $C_{e^-}=\alpha E /(Q^4 E^\prime)$ where $E$ is the energy of the initial lepton, $\alpha=1/137$ is the electromagnetic fine structure constant, $Q^2={\bf q}^2-\omega^2$ is the four-momentum transfer
and the leptonic tensor is
\begin{align}
L_{\mu \nu}  = \frac{1}{EE^\prime} (k_\mu k^\prime_\nu + k^\prime_\mu k_\nu - g_{\mu\nu}\, (k \cdot k^\prime - m_e^2))\, ,
\end{align}
where $m_e=511$ keV is the electron mass. { Note that we adopted the convention $h=c=1$.} For CC electroweak interactions, we have that a neutrino or anti-neutrino scatters off the initial nucleus and in the final state the corresponding charge lepton is emitted. The prefactor reads $C_l=(G_F \cos\theta_c)^2/(4\pi) |{\bf k}^\prime| E^\prime $ with $G_F=1.1803 \times 10^{-5}$~\cite{Nakamura:2002jg} and $\cos\theta_c=0.97425$~\cite{ParticleDataGroup:2010dbb}. The leptonic tensor has an additional term proportional to the Levi-Civita tensor
\begin{align}
L_{\mu \nu}  = \frac{1}{EE^\prime} (k_\mu k^\prime_\nu + k^\prime_\mu k_\nu - g_{\mu\nu}\, k \cdot k^\prime \pm \epsilon_{\mu\rho\nu\sigma}k^\rho {k^\prime}^\sigma)\, ,
\end{align}
where the sign +(-) corresponds to a $\nu$ ($\bar{\nu}$) in the initial state. 

The hadronic response tensor, $R^{\mu\nu}$, contains all the information on the structure of the nuclear target and is defined as
\begin{equation}\label{eq:ResponseFunctionE}
    R_{\mu\nu} = \sum_{f}\langle\Psi_{0}|J^{\dagger}_\mu|\Psi_{f}\rangle\langle \Psi_{f}|J_{\nu}|\Psi_{0}\rangle\delta (E_{0} + \omega - E_{f})\, ,
\end{equation}
in terms of a sum over all transitions from the ground state $|\Psi_{0}\rangle$ with energy $E_0$ to any final state $|\Psi_{f}\rangle$ with energy $E_f$, including states with additional hadrons. The nuclear current operator describing the interaction with the electroweak probe is denoted by $J_\mu$.

\subsection{Nuclear Hamiltonian and current operator}
\label{sec:ham_curr}
Microscopic nuclear methods are aimed at describing properties of nuclear systems as they emerge from the individual interactions among the constituent protons and neutrons. This endeavor is based on the tenet that the internal structure of atomic nuclei can be described starting from a non-relativistic Hamiltonian of $A$ point-like nucleons
\begin{equation}\label{eq:hamiltonian}
    H = \sum_{i}^{A}\frac{\mathbf{p}_{i}^{2}}{2m_{N}} + \sum_{i<j}^{A}v_{ij} + \sum_{i<j<k}^{A}V_{ijk}\, .
\end{equation}
In the above equation $\mathbf{p}$ and $m_{N}$ are the nucleon momentum and mass defined as the average of the proton and neutron mass $m_{N}=(m_p+m_n)/2$, while $v_{ij}$ and $V_{ijk}$ are the two (NN) and three-nucleon (3N) potentials respectively; four- and higher-body potentials are assumed to be suppressed.

Phenomenological NN interactions have been traditionally constructed by including the long-range one-pion exchange interaction, while different schemes are implemented to account for intermediate and short range effects, including multiple-pion-exchange, contact terms, heavy-meson-exchange, or excitation of nucleons into virtual $\Delta$-isobars. As an example, the highly-accurate Argonne $v_{18}$ (AV18) potential~\cite{Wiringa:1994wb}, involves a number of parameters that are determined by 
fitting deuteron properties and the large database of NN scattering data at laboratory energies up to pion production threshold. The AV18 potential is written as
\begin{equation}
    v_{ij} = \sum_{p=1}^{18} v_{p}(r_{ij})O^{p}_{ij}\, .
    \label{eq:NN_pot}
\end{equation}
The first 14 spin-isospin operators are charge independent {
\begin{equation}
O^{1-14}_{ij} = [1, \boldsymbol{\sigma}_i \cdot \boldsymbol{\sigma}_j, S_{ij}, \mathbf{L}\cdot \mathbf{S}, \mathbf{L}^2, \mathbf{L}^2 (\boldsymbol{\sigma}_i \cdot \boldsymbol{\sigma}_j), (\mathbf{L}\cdot \mathbf{S})^2] ] \times [1, \boldsymbol{\tau}_i \cdot \boldsymbol{\tau}_j]\, ,
\label{eq_NN_pot_1_14}
\end{equation}
where $\boldsymbol{\sigma}_i$ are Pauli matrices that operate over the spin of nucleons, $S_{ij} = 3 (\hat{r}_{ij} \cdot \boldsymbol{\sigma}_i) (\hat{r}_{ij} \cdot \boldsymbol{\sigma}_j) - \boldsymbol{\sigma}_i \cdot \boldsymbol{\sigma}_j$ is the tensor operator, $\mathbf{L}_{ij}=\frac{1}{2i}(\mathbf{r}_i - \mathbf{r}_j) \times (\boldsymbol{\nabla}_i - \boldsymbol{\nabla}_j)$ is the relative angular momentum of the pair $ij$ and $\mathbf{S} = \frac{1}{2}(\boldsymbol{\nabla}_i+ \boldsymbol{\nabla}_j)$ is the total spin. The remaining operators include three charge-dependent terms and one charge-symmetry breaking contribution
\begin{align}
O^{15-17}_{ij} &= [1, \boldsymbol{\sigma}_i \cdot \boldsymbol{\sigma}_j, S_{ij} ] \times T_{ij}\, , \nonumber\\
O^{18}_{ij} &=  \tau_i^z + \tau_j^z\, ,
\end{align}
where $T_{ij} = 3 \tau_i^z \tau_j^z - \boldsymbol{\tau}_i \cdot \boldsymbol{\tau}_j$ is the isotensor operator.  }

Phenomenological 3N interactions, consistent with the NN ones, are generally expressed as a sum of a two-pion-exchange P-wave term, a two-pion-exchange S-wave contribution, a three-pion-exchange contribution, plus a contact interaction. Their inclusion is essential for reproducing the energy spectrum of atomic nuclei and saturation properties of infinite nucleonic matter. For instance, the Illinois-7 3N force~\cite{Pieper:2008}, when used together with AV18, can reproduce the spectrum of nuclei up to $^{12}$C with percent-level accuracy ---- see Fig.~\ref{fig:nuclearaccuracy} discussed in Sec.~\ref{sec:qmc_approaches}. 

The past two decades have witnessed the tremendous development and success of chiral Effective Field Theory~\cite{Weinberg:1990rz,Weinberg:1991um,VanKolck:1993ee,Ordonez:1992xp,Ordonez:1995rz,Bernard:1995dp,Epelbaum:2008ga,Epelbaum:2012zz,Epelbaum:2014efa,Entem:2003ft,Machleidt:2011zz,ekstrom2015accurate} ($\chi$EFTs). This formalism exploits the broken chiral symmetry pattern of QCD, the fundamental theory of strong interactions, to construct an effective Hamiltonian organized in powers the ratio between the pion mass, $m_\pi$, or a typical nucleon momentum, $Q$, and the scale of chiral symmetry breaking, $\Lambda_\chi \sim 1$ GeV. Over the years, NN interactions have been developed up to $\mathrm{N}^{5}\mathrm{LO}$ in the chiral expansion~\cite{Entem:2015xwa,Epelbaum:2015pfa,Reinert:2017usi}, with a full systematic error analysis currently underway~\cite{Wesolowski:2021cni}. On the other hand, chiral 3N forces have been fully derived at $\mathrm{N}^{3}\mathrm{LO}$, while only contact terms at $\mathrm{N}^{4}\mathrm{LO}$ have so far been included~\cite{Girlanda:2023znc}.

In analogy with the nuclear Hamiltonian, the nuclear current operator $J^{\mu}$, which couples the nucleus to the external electroweak probe,
can be written as a sum of both one and two-body contributions
\begin{equation}
    J^{\mu}_{A}(q) = \sum_{i}j^{\mu}_{i}(q) + \sum_{ij}j^{\mu}_{ij}(q) + ...\,
\end{equation}
where higher order terms, involving three nucleons or more, are found to be small ~\cite{Marcucci:2005zc} and generally neglected.

The one-body electromagnetic current is given by
\begin{align}
j^\mu_{\rm EM}(q)= j^\mu_{\gamma, S}(q)+j^\mu_{\gamma, z}(q)\, ,
\end{align}
where the first term is the isoscalar contribution and the second one is the isovector. The isoscalar component reads
\begin{align}
j^\mu_{\gamma, S}(q)= \frac{G_E^{S}+\tau G_M^{S}}{2(1+Q^2/4m_N^2)} \gamma^\mu+ i  \frac{\sigma^{\mu\nu}q_\nu}{4m_N}\frac{G_M^{S}-G_E^{S}}{1+Q^2/4m_N^2}
\, .
\label{rel:1b:curr}
\end{align}
The isoscalar and isovector component of electric and magnetic form factors are written in terms of the proton and neutron ones as
\begin{align}
G_{E,M}^S=G_{E,M}^p+G_{E,M}^n,\ \ \ \ \ \ \ \ G_{E,M}^V=G_{E,M}^p-G_{E,M}^n\, .
\end{align}
The isovector contribution to the current operator $j^\mu_{\gamma,z}$ is obtained by replacing $G_{E,M}^S\rightarrow G_{E,M}^V \tau_z$ in Eq.~\eqref{rel:1b:curr}.

\begin{figure}[h!]
    \includegraphics[width=11.5cm]{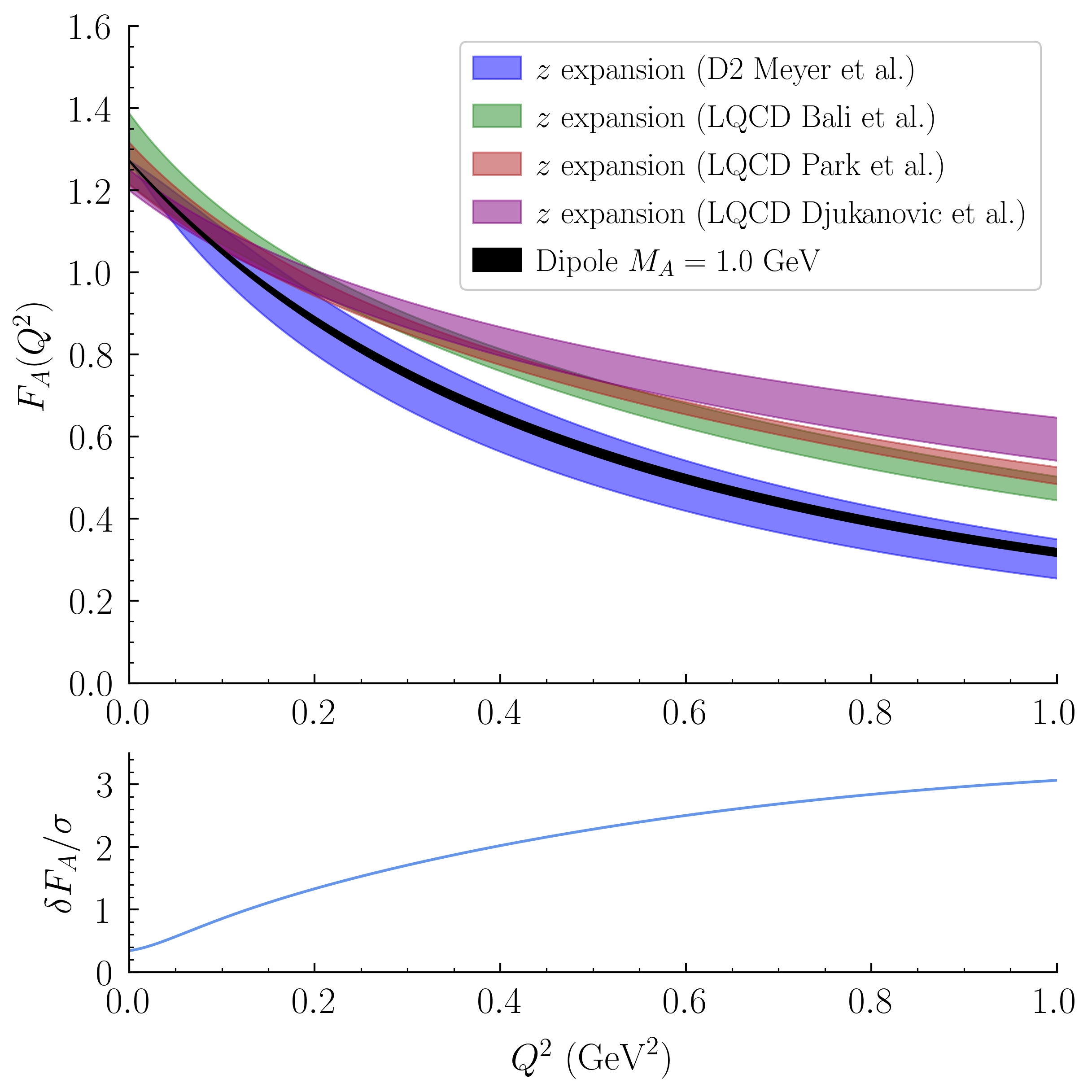} 
    \caption{The blue band in the top panel displays the  
    nucleon axial form factor determined using fits to neutrino-deuteron scattering data with the model-independent $z$ expansion from Ref.~\cite{Meyer:2016oeg} (D2 Meyer et al.). LQCD results are shown for comparison from Ref.~\cite{RQCD:2019jai} (LQCD Bali et al., green), Ref.~\cite{Park:2021ypf} (LQCD Park et al., red) and Ref.~\cite{Djukanovic:2022wru} (LQCD Djukanovic et al., purple). Bands show combined statistical and systematic uncertainties in all cases. A dipole parameterization with $\Lambda_A = 1.0$ GeV and a $1.4\%$ uncertainty~\cite{Bodek:2007ym} is also shown for comparison (black). The lower panel shows the absolute value of the difference between D2 Meyer et al. and LQCD Bali et al. results divided by their uncertainties added in quadrature; very similar results are obtained using the other LQCD results. Figure from Ref.~\cite{Simons:2022ltq}.}
    \label{fig:form_factors}
\end{figure}

The one-body charge and current operator employed in the GFMC are obtained from the nonrelativistic reduction of the covariant operator of Eq.~\eqref{rel:1b:curr} including all the terms up to $1/m_N^2$. This expansion leads to the following expressions for isoscalar charge, 
 transverse ($\perp$) and longitudinal ($\parallel$) to ${\bf q}$ components of the current operator  
\begin{align}
j^0_{\gamma,S}=& \frac{G_E^S}{2\sqrt{1+Q^2/4m_N^2}}-i\frac{2 G_M^S-G_E^S}{8m^2_N}{\bf q}\cdot (\pmb{\sigma}\times {\bf p})\nonumber\\
{\bf j}^\perp_{\gamma,S}=&\frac{G_E^S}{2m_N}{\bf p}^\perp-i\frac{G_M^S}{4m_N}({\bf q}\times{\pmb \sigma})\nonumber\\
j^\parallel_{\gamma,S}=&\frac{\omega}{|{\bf q}|}j^0_{\gamma,S}\, .
\end{align}
Note that the last relation has been obtained from the conserved vector current (CVC) relation~\cite{Gell-Mann:1960eja}, e.g. $\omega J^{0}(\omega,{\bf q}) - {\bf q} \cdot \mathbf{J}(\omega,{\bf q}) = 0$.
The CC electroweak interactions of a neutrino or anti-neutrino with the hadronic target are written as the sum of a vector and axial term
 \begin{align}
 j^\mu_{CC}(q)=&j^\mu_{V,\pm}(q)+j^\mu_{A,\pm}(q)\nonumber\\ .
  \end{align}
The CVC hypothesis allows one to write $j^\mu_{V,\pm}(q)$ in terms of the isovector term where $\tau_{z}$ is replaced by the isospin raising-lowering operator $\tau_{\pm}=(\tau_{i,x}\pm \tau_{i,y})/2$. The relativistic expression of the axial one-body current operator reads
\begin{align}
 j^\mu_{A, \pm}=-\gamma^\mu \gamma_5 G_A \tau_\pm -q^\mu \gamma_5 \frac{G_P}{m_N}\tau_\pm\ .
\end{align}
Based on Partially Conserved Axial Current (PCAC) arguments, the pseudo-scalar form factor is written in terms of the axial one 
\begin{align}
G_P=\frac{2m_N^2}{(m_\pi^2+Q^2)}G_A\, . 
\end{align}
Most neutrino-nucleus scattering calculations are carried out employing a dipole parameterization for the axial form factor, which is given by 
\begin{align}
G_A &=\frac{g_A}{( 1+ Q^2/ \Lambda_A^2 )^2}\ ,
\end{align}
where the nucleon axial-vector coupling constant is taken to be $g_A=1.2694$~\cite{ParticleDataGroup:2016lqr} and the axial mass is taken as $\Lambda_{A} = 1.049$ GeV~\cite{Nieves:2011yp}. More recently, a model-independent $z$ expansion has been introduced to parameterize the axial form factor 
\begin{equation}
  G_{A}(Q^2) = \sum_{j=0}^{\infty} a_{j} \, z(Q^2)^{j} \approx \sum_{j=0}^{j_{\rm max}} a_{j} \, z(Q^2)^{j}\, .
  \label{eq:zexp}
\end{equation}
{ In the last equation, $z$ is an analytic function of $Q^2$ for $Q^2 = -t > -t_c$
\begin{equation}
  z(Q^2) = \frac{\sqrt{t_c + Q^2} - \sqrt{t_c - t_0}}{\sqrt{t_c + Q^2} + \sqrt{t_c - t_0}},
  \label{eq:z}
\end{equation}
where $t_c$ is the location of the $t$-channel cut~\cite{Hill:2006ub,Hill:2010yb,Bhattacharya:2011ah}
and $t_0$ is an arbitrary parameter.} The coefficients $a_j$ include nucleon structure information and $j_{\rm max}$ is a truncation parameter required to make the number of expansion parameters finite. The coefficients of this expansion are determined by fitting either neutrino-deuteron scattering data~\cite{Meyer:2016oeg} or Lattice-QCD nucleon axial-current matrix elements at several discrete values of $Q^2$ ~\cite{RQCD:2019jai,Park:2021ypf,Djukanovic:2022wru}. The results of these different determinations of the axial form factors are displayed in Fig.~\ref{fig:form_factors}. While an agreement between different LQCD calculations is clearly visible, the LQCD axial form factor results are 2-3$\sigma$ larger than the results of Ref.~\cite{Meyer:2016oeg} for $Q^2 \gtrsim 0.3\text{ GeV}^2$. The impact of these tensions in the  $Q^2$ dependence of the axial form factor on neutrino-nucleus cross-section predictions 
has been discussed in Ref.~\cite{MiniBooNE:2010bsu,Bernard:2001rs} and recently in \cite{Meyer:2022mix,Simons:2022ltq}.


For the CC processes, we report the nonrelativistic reduction of the charge and axial current operators~\cite{Carlson:1997qn} (for brevity we neglect order $1/m_N^2$ terms)
\begin{align}
j^0_{A,\pm}=-\frac{G_A}{2m_N}\tau_{\pm}{\bm \sigma}\cdot({\bf q}+{\bf p})\, ,\ \ \ \ \ \ \ {\bf j}_{A,\pm}=-\frac{G_A}{2m_N}{\bm \sigma}\tau_{\pm}\ .
\label{axial:curr:nr}
\end{align}
and the pseudoscalar contribution 
\begin{align}
j^\mu_{PS,\pm}=\frac{G_A}{m_\pi^2+Q^2}\tau_{\pm}q^\mu{\bm \sigma}\cdot{\bf q}\, .
\end{align}

The current conservation relation can be rewritten as 
\begin{equation}
    \mathbf{\nabla}\cdot\mathbf{J}_{\mathrm{EM}} + i[H,J^{0}_{\mathrm{EM}}] = 0\, .
\end{equation}
{ It requires the introduction of a two-body current operator in $J^\mu_{\rm EM}$ and links the divergence of this operator to the commutator of the charge operator with the nucleon-nucleon interaction.}

For the electromagnetic case, however, gauge invariance actually puts constraints on these form factors by linking the divergence of the two-body currents to the commutator of the charge op- erator with the nucleon-nucleon interaction

Within $\chi$EFT one can exploit the gauge invariance of the theory and construct nuclear current operators that are fully consistent with the nuclear potentials, at each order of the chiral expansion. The derivation of $\chi$EFT two-body electroweak currents has been the subject of extensive study carried out by different groups~\cite{Pastore:2008ui,Pastore:2009is,Pastore:2011ip,Kolling:2009iq,Kolling:2011mt,Baroni:2017gtk,Baroni:2018fdn}. 
 
The majority of the results that will be presented in this review has been obtained utilizing semi-phenomenological currents that are consistent with the AV18 potential. 
The isoscalar and isovector components of the two-body electromagnetic current operator consist of ``model-independent'' and ``model-dependent term'' terms. 
The former are obtained from the NN interaction, and by construction satisfy current conservation. They consists of the one-pion and one-rho exchange current operator --- their expressions are well known and reported in 
Refs.~\cite{Dekker:1994yc,Rocco:2018mwt,Schiavilla:1989zz} both in their relativistic and non relativistic formulation. \\
The transverse components of the two-body currents cannot be directly linked to the nuclear Hamiltonian. 
The isovector current is associated with the exchange of a pion followed by the excitation of a $\Delta$-resonance in the intermediate state.
The isoscalar contribution includes the $\rho\pi\gamma$ transition whose couplings are extracted from the widths of the radiative decay $\rho \rightarrow \pi\gamma$ and the $Q^2$ dependence of the electromagnetic transition form factor is modeled assuming vector-meson dominance~\cite{Berg:1980lwp,RevModPhys.70.743}.

\section{Quantum Monte Carlo Approaches}
\label{sec:qmc_approaches}

Solving the Schr\"odinger equation for the nuclear Hamiltonian defined in Eq.~\eqref{eq:hamiltonian} entails nontrivial difficulties, owing to the nonperturbative nature and strong spin-isospin dependence of realistic nuclear forces. The VMC method is routinely employed to approximately find the ground-state solution of the quantum many-body problem for nuclei with up to $A=12$ nucleons~\cite{Carlson:2014vla}. Within this approach, the true ground state $\Psi_0$ is approximated by a variational state $\Psi_{V}$, which is defined in terms of a set of variational parameters. The optimal values of the latter are found exploiting the variational principle, i.e. by minimizing the variational energy
\begin{equation}
E_{V} = \frac{\langle\Psi_{V}|H|\Psi_{V}\rangle}{\langle \Psi_{V}|\Psi_{V}\rangle} \geq E_{0}\, .
\end{equation}
The form of the variational state is taken to be
\begin{equation}
    \Psi_{V} = \mathcal{F} |\Phi\rangle,
\end{equation}
where $\mathcal{F}$ is a permutation-invariant correlation operator of a Jastrow, and the anti-symmetric $|\Phi\rangle$ controls the quantum numbers and the long-range behavior of the wave function. { The correlation operator explicitly includes correlations between pairs and triplets of nucleons
\begin{equation}
    \mathcal{F}=\Big(\mathcal{S} \prod_{i<j<k} (1 + F_{ijk}) \Big)\Big(\mathcal{S} \prod_{i<j} F_{ij} \Big) \, ,
\end{equation}
where $S$ is the symmetrization operator, which is required to ensure the anti-symmetry of $\Psi_{V}$ since, in general, neither the two-body correlations, $F_{ij}$, nor the three-body ones, $F_{ijk}$, commute.} The structure of the spin-dependent nuclear correlation operators reflects the one of the NN potential of Eq.~\eqref{eq:NN_pot}
\begin{equation}
F_{ij}=\sum_{p=1}^6 u^p(r_{ij})O^p_{ij}\, ,
\end{equation}
{ where the first six operators of Eq.~\eqref{eq_NN_pot_1_14} are $O^{1-6}_{ij} = [1, \boldsymbol{\sigma}_i \cdot \boldsymbol{\sigma}_j, S_{ij}, ] \times [1, \boldsymbol{\tau}_i \cdot \boldsymbol{\tau}_j]$. } More sophisticated correlation operators that explicitly include spin-orbit correlations have been used in the cluster variational Monte Carlo calculations of Ref.~\cite{Lonardoni:2017egu}. However, the computational cost of these additional terms is significant, while the the gain in the variational energy is relatively small~\cite{Wiringa:2000gb}.

\begin{figure}[h]
    \includegraphics[width=13cm]{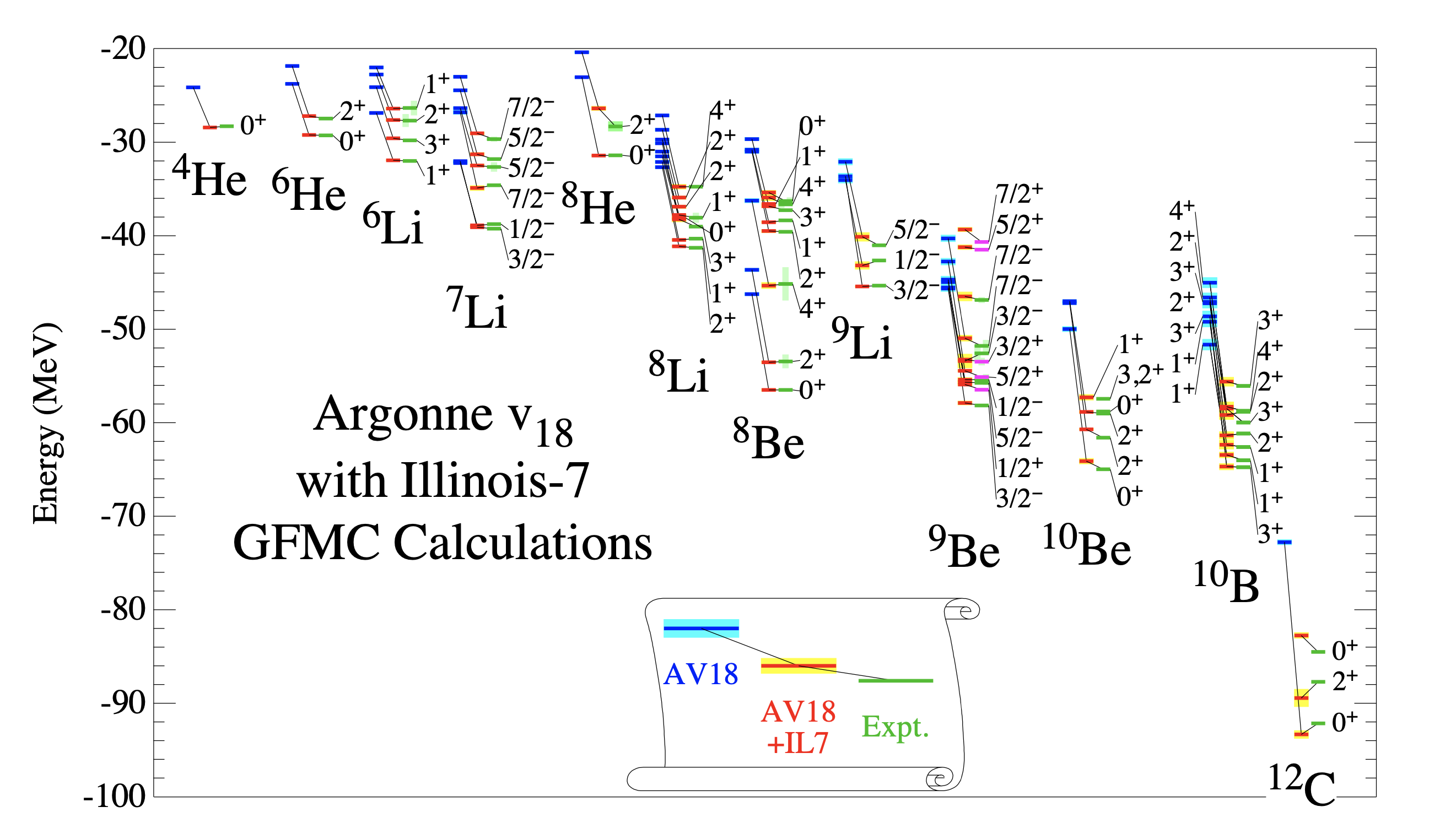} 
    \caption{Energies of light nuclear ground and excited states from a particular parameterization of Eq.~\eqref{eq:hamiltonian} computed using Green's Function Monte Carlo (GFMC) techniques. Figure from~\cite{Carlson:2014vla}.}
    \label{fig:nuclearaccuracy}
\end{figure}

The GFMC evolves the variational state in imaginary time to filter out the excited state components, so that 
\begin{equation}
 |\Psi_0\rangle = \lim_{\tau\to\infty} |\Psi(\tau)\rangle = \lim_{\tau\to\infty}{\rm exp}[-(H-E_0)\,\tau]\,|\Psi_T\rangle\, ,
 \label{eq:psi:gs}
\end{equation}
The above imaginary-time evolution is carried out as a series of many small steps $\Delta\tau$ using an exact two-body short-time propagator~\cite{Pudliner:1997ck}. At each step, the GFMC retains all of the  spin-isospin components of the nuclear wave function and can take as input the most realistic local interactions. The results for the ground state energies of nuclei up to $^{12}$C has been computed with 1\% accuracy within GFMC using the semi-phenomenological AV18+IL7 potentials in Ref.~\cite{Carlson:2014vla} and they are displayed in Fig.~\ref{fig:nuclearaccuracy}. Note that a plot with a comparable degree of accuracy has been also obtained using as input the $\Delta$-full $\chi$EFT nuclear forces that are local in coordinate space~\cite{Piarulli:2015mra,Piarulli:2016vel}.

Since all the spin-isospin degrees of freedom are retained, the GFMC suffers from an exponential scaling with the number of nucleons, which currently limits its applicability to light nuclei, up to $^{12}$C. The Auxiliary Field Diffusion Monte Carlo can reach larger nuclear systems by representing the spin-isospin degrees of freedom in terms of products of single-particle states, thereby reducing the computational cost from exponential to polynomial in $A$~\cite{Carlson:2014vla,Schmidt:1999lik}. However, the use of Hubbard-Stratonovich transformations in the AFDMC imaginary-time propagation prevents the AFDMC from treating highly-realistic NN potentials that include an isospin-dependent spin-orbit term.

\subsection{Green's Function Monte Carlo calculations of electroweak responses}

GFMC techniques go beyond just the calculation of ground state energies and wave functions. Dynamical properties of the nucleus can be extracted by reducing the sum over final states in Eq.~\eqref{eq:ResponseFunctionE} to the expectation value of a kernel operator evaluated in the ground state. More specifically, we consider the Euclidean response function 
\begin{equation}
    \begin{aligned}
    E_{\alpha\beta}(\mathbf{q},\tau) &= \int d\omega K(\tau, \omega) R_{\alpha\beta}(\mathbf{q},\omega)\nonumber\\
    &= \sum_{f}\langle\Psi_{0}|J^{\dagger}_{\alpha}(\mathbf{q})|\Psi_{f}\rangle K(\tau, E_{f} - E_{0})\langle\Psi_{f}|J_{\beta}(\mathbf{q})|\Psi_{0}\rangle\, ,
    \end{aligned}
\end{equation}
where $K(\tau,\omega)$ is a yet to be specified kernel. Using a completeness relation amongst the final states this can be simplified to
\begin{equation}
    E_{\alpha\beta}(\mathbf{q},\tau) = \sum_{f}\langle\Psi_{0}|J^{\dagger}_{\alpha}(\mathbf{q}) K(\tau, H - E_{0})J_{\beta}(\mathbf{q})|\Psi_{0}\rangle\, ,
\end{equation}
so that the problem involves only the ground state. Choosing an appropriate kernel function allows one to solve for the Euclidean response using \textit{ab initio} methods. In particular, a Laplace kernel has been adopted with GFMC techniques yielding the following expression for the inelastic contribution to the response function is
{ 
\begin{align}
E_{\alpha\beta}({\bf q},\tau) &= \int_{0}^{\infty} d\omega R_{\alpha\beta}e^{-\omega\tau}
= \langle \Psi_0 | J_\alpha^\dagger ({\bf q}) e^{-(H-E_0)\tau} J_\beta({\bf q}) | \Psi_0\rangle\, .
\end{align}
In the electromagnetic case, only the longitudinal $(R_L = R_{00})$ and transverse $(R_T = R_{xx}+R_{yy})$ responses contribute. In the longitudinal case, we remove the elastic contribution, in which the final state is simply the recoiling ground state, by defining
\begin{align}
E_{00}({\bf q},\tau) &= \int_{\omega^{+}_{el}}^{\infty} d\omega R_{00}e^{-\omega\tau}
= \langle \Psi_0 | J_0^\dagger ({\bf q}) e^{-(H-E_0)\tau} J_0({\bf q}) | \Psi_0\rangle - |F_0(\mathbf{q})|^2 e^{-\omega_{el}\tau}\, ,
\label{eq:eucl_elastic}
\end{align}
In the above equation, $\omega_{el}= \mathbf{q}^2/2M_A$,  with $M_A$ being the mass of the nucleus, is the energy of the recoiling ground state and} the elastic form factor is defined as $F_0(\mathbf{q}) = \langle \Psi_0 | J_0({\bf q})| \Psi_0 \rangle$. 

The calculation of the imaginary-time correlation operator in the right hand side of Eq.~\eqref{eq:eucl_elastic} follows the same methodology applied to project out the exact ground state of $H$ from a trial wave function in Eq.~\eqref{eq:psi:gs}. First, an unconstrained imaginary-time propagation of the state $|\Psi_0\rangle$ is performed and stored. Then, the states $J_\alpha({\bf q})|\Psi_0\rangle$ are evolved in imaginary time following the path previously saved. For a complete discussion of the methods see Refs.~\cite{Lovato:2015qka,Lovato:2016gkq,Lovato:2017cux}. 
To retrieve the energy dependence of the response functions Bayesian techniques, most notably maximum-entropy (MaxEnt), have been developed specifically for this type of problem~\cite{Lovato:2015qka} and successfully exploited to obtain the smooth quasi-elastic responses~\cite{Lovato:2016gkq,Lovato:2017cux}. However, MaxEnt struggles to reconstruct the narrow peaks corresponding to low-energy transitions. In particular, understanding the low-lying nuclear transitions is necessary to  properly describe the longitudinal electromagnetic responses of $^{12}$C in the low-energy transfer. The results of Ref.~\cite{Lovato:2016gkq} have been obtained by subtracting the contribution of these excited states { by defining 
\begin{align}
\bar{E}_{00}({\bf q},\tau) = E_{00}({\bf q},\tau) - \sum_f \left| \langle \Psi_f| J_0(\mathbf{q}) | \Psi_0 \rangle \right|^2 e^{-(E_f - E_0)\tau}\, ,
\end{align}
where the sum only includes the $2^+$, $0_2^+$, and $4^+$. 
final states. The experimental energies and longitudinal transition form factors from Refs.~\cite{Bryan:1971nz,Chernykh:2010zu} are used.}
 
\begin{figure}[h!]
    \centering
    \includegraphics[width=0.48\linewidth]{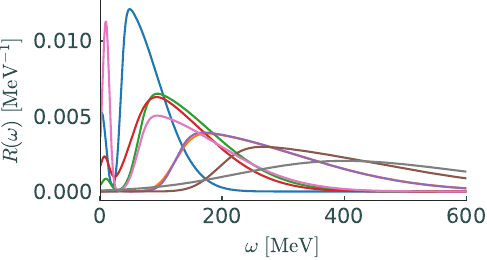}
    \hspace{0.15cm}
    \includegraphics[width=0.48\linewidth]{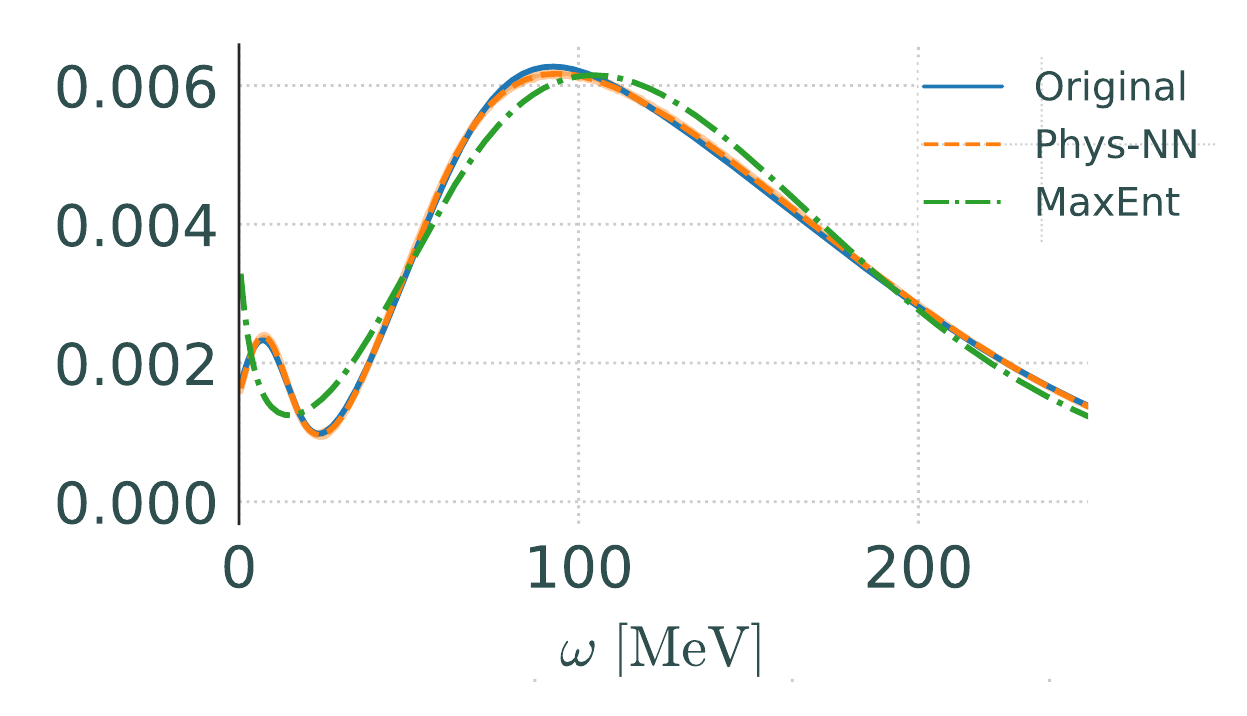}
    \caption{Left Panel: Training data examples of response functions characterized by an elastic narrow peak in addition to the quasi-elastic peak~\cite{Raghavan:2020bze}. Right Panel: Comparison between the Phys-NN and MaxEnt reconstructions for the one-peak data set, adapted from Ref.~\cite{Raghavan:2020bze}.}
    \label{fig:GFMC-resp}
\end{figure}

Furthermore, understanding this region is also crucial to detect supernova neutrinos as well as to describe the low-energy tail contribution of the neutrino flux in accelerator experiments. To this aim, in Ref.~\cite{Raghavan:2020bze} an exploratory study has been carried out to develop physics-informed artificial neural network architectures suitable for approximating the inverse of the Laplace transform, utilizing simulated, albeit realistic, electromagnetic response functions. The training has been performed using pairs of physically meaningful responses and their Laplace transform. There are two data sets of response functions, characterized by either one or two distinct peaks in the energy-transfer domain. The left panel of Figure~\ref{fig:GFMC-resp} displays a subset of the two-peaks training data.
A detailed comparison of the reconstruction results obtained for both the one- and two-peak data sets demonstrates that the physics-informed artificial neural network outperforms MaxEnt in both the low-energy transfer and the quasi-elastic regions --- an illustrative example of this trend is shown in the right panel of Figure~\ref{fig:GFMC-resp}. Work is currently underway to extend the study of Ref.~\cite{Raghavan:2020bze} to real GFMC data and to perform error propagation.

\subsubsection{Relativistic Corrections}\label{sec:relcor}
One of the limitations of the GFMC approach to describe nuclear reactions is the nonrelativistic formulation of the many-body problem. Although the leading relativistic corrections are typically included in the transition operators~\cite{Shen:2012xz}, the kinematics of the reaction is treated as nonrelativistic, and an expansion of fully relativistic currents in $p/m$ is made. The explicit expression of the one-body current operators adopted in the GFMC calculation are reported in Sec.~\ref{sec:ham_curr}. Thereby the application of these methods is limited to moderate values of the momentum transfer. 

In a number of works~\cite{Efros:2005uk,Efros:2009qp,Yuan:2010gh,Efros:2010fe,Yuan:2011rd,Rocco:2018,Nikolakopoulos:2023zse}, a method was proposed to
extend the applicability of manifestly nonrelativistic hyperspherical-harmonics and Quantum Monte Carlo (QMC) methods to higher momentum transfer values than typically possible. This method reduces relativistic effects by performing the calculations in a reference frame that minimizes nucleon momenta. 
The reference frame that achieves this goal for kinematics close to the quasi-elastic peak is the active nucleon Breit frame (ANB).
The ANB is defined as the reference frame moving along the direction of the momentum transfer $\mathbf{q}$ where $\mathbf{P}_i^{ANB} = -A \mathbf{q}^{ANB}/2$, with $\mathbf{P}_i^{ANB}$ the momentum of the initial nucleus in the ANB.
Indeed, if one assumes that the bulk of the momentum is transferred to a single nucleon, in the ANB this nucleon has initial momentum $\mathbf{k} \approx \mathbf{P}^{ANB}_0/A = -\mathbf{q}^{ANB}/2$. The corresponding final-state nucleon has momentum $\mathbf{k} + \mathbf{q}^{ANB} = \mathbf{q}^{ANB}/2$. Hence in the ANB the magnitude of both the initial and final-state nucleon momentum is minimal. Additionally, the energy transfer at the quasi-elastic peak is zero in the ANB frame, implying that $\mathbf{q}^{ANB}$ is also minimal at the quasi-elastic peak compared to other frames.

Within a non-relativistic calculation, the nuclear response can be computed in different reference frames by evaluating  Eq.(\ref{eq:ResponseFunctionE}) at the momentum transfer in the reference frame specified by ${\bf q}^{fr}$, and by taking into account the kinetic energy of initial- and final-state systems in the energy balance. Thus the energy-conserving delta function $\delta( E_0 - \omega - E_f)$ is evaluated with $\omega = \omega^{fr}$, and $E_0 - E_f = -(\Pv^{fr}_f)^2/(2M_A)+(\Pv^{fr}_0)^2/(2M_A)$, leading to an energy-shift of the response.

The dependence on the reference frame used for calculations can be evaluated by performing a Lorentz boost of the response back to the LAB frame. At momentum transfers larger than 500 $\mathrm{MeV}$ one starts to see differences between calculations performed in different reference frames~\cite{Efros:2005uk,Rocco:2018,Nikolakopoulos:2023zse}, indicating that relativistic effects become important.

This frame dependence in the region of the quasi-elastic peak can be significantly reduced by including the assumption of single nucleon knockout in the energy balance. 
In order to achieve this, one can use the so-called two-fragment model, where a breakup into two fragments, the nucleon and residual system is assumed. Following the arguments of Refs.~\cite{Efros:2005uk,Rocco:2018}, the approach consists of evaluating the nuclear response at an energy $p_{rel}^2/(2\mu)$ with $p_{rel}$ the magnitude of the relative momentum of the two fragments and $\mu$ the reduced mass.
The energy of the final-state system can be written in a relativistic way as
\begin{align}
\label{rel_kin}\nonumber
\omega + E_0 = E_f &= \sqrt{m^2 + ({\bf p}_{rel} + (\mu/M){\bf P}_f)^2} \\
         &+  \sqrt{M^2 + ({\bf p}_{rel} - (\mu/m){\bf P}_f)^2} \,;
\end{align}
where $\mathbf{P}_f = \mathbf{P}_0 + \mathbf{q}$ is center-of-mass momentum. Under the assumption that $\mathbf{p}_{rel}$ is directed along $\mathbf{q}$ one can solve Eq.~(\ref{rel_kin}) for $p_{rel}$.

In Refs.~\cite{Efros:2005uk,Rocco:2018} it is indeed found that the frame dependence for electroweak scattering is strongly reduced when including the two-fragment model to determine the energy. Moreover the resulting LAB frame responses are practically identical to the response obtained in the ANB when the fragment model is not included~\cite{Nikolakopoulos:2023zse}. 

Calculations of the nuclear response in the ANB can be used to extend the applicability of GFMC responses to larger momentum transfer. In Ref.~\cite{Rocco:2018} an improved description of $(e,e')$ data for scattering off ${}^{4}$He was obtained at large momentum transfer with GFMC responses computed in the ANB. Recently this approach was applied to GFMC calculations of flux-folded charged-current neutrino scattering off ${}^{12}$C~\cite{Nikolakopoulos:2023zse}.

\section{Extended Factorization Scheme}\label{sec:fact}
At large values of the momentum transfer, ($|\textbf{q}| \gtrsim 400$ MeV), the Impulse Approximation (IA) can be applied in which the lepton-nucleus scattering is approximated as an incoherent sum of scatterings with individual nucleons, and the struck nucleon system is decoupled from the rest of the final state spectator system. 
\subsection{One Body Currents}
We begin with retaining only one body current terms and factorize the final state according to
\begin{equation}
    |\Psi_{f}\rangle = |\textbf{p}'\rangle \otimes |\Psi^{A-1}_{f},\textbf{p}_{A-1}\rangle\, ,
    \label{eq:fact}
\end{equation}
where $|\textbf{p}'\rangle$ is the final state nucleon produced at the vertex, assumed to be in a plane wave state and on-shell, and $|\Psi^{A-1}_{f},\textbf{p}_{A-1}\rangle$ describes the residual system, carrying momentum $\textbf{p}_{A-1}$. Inserting this factorization ansatz as well as a single-nucleon completeness relation gives the matrix element of the one body current operator as
\begin{equation}\label{eq:1bodyME}
    \langle \Psi_{f}| j^{\mu} |\Psi_{0}\rangle \rightarrow \sum_{k}[\langle \Psi^{A-1}_{f}|\otimes\langle k|] |\Psi_{0}\rangle\langle p|\sum_{i}j^{\mu}_{i}|k\rangle\, ,
\end{equation}
where $\textbf{p} = \textbf{q} + \textbf{k}$. This first piece of the matrix element explicitly does not depend on the momentum transfer and so can be computed using techniques in nuclear many body theory. The second piece can be straightforwardly computed once the currents $j^{\mu}_{i}$ are specified as the single nucleon states are just free Dirac spinors. { It is important to point out that factorization allows for an account of relativistic effect by adopting Dirac quadri-spinors for the description of the struck particles in the initial and final states and the current operator of Eq.~\eqref{rel:1b:curr}. These effects become extremely important at large values of $\textbf{q}$ and $\omega$ where a non-relativistic calculation is no longer reliable.} Substituting the last equation into Eq.~\eqref{eq:ResponseFunctionE}, and exploiting momentum conservation at the single nucleon vertex, allows us to rewrite the incoherent contribution to the one body hadron tensor as
\begin{equation}\label{eq:W1body}
\begin{aligned}
    R_{1b}^{\mu\nu}(\textbf{q},\omega) = &\int\frac{d^{3}k}{(2\pi)^3}dEP_{h}(\textbf{k},E)\frac{m^{2}_{N}}{e(\textbf{k})e(\textbf{k} + \textbf{q})} \times\sum_{i}\langle k|j_{i}^{\mu\dagger}|k +q\rangle\langle k+q|j_{i}^{\nu}|k\rangle \\ 
    & \times\delta(\tilde{\omega} + e(\textbf{k}) - e(\textbf{p}))\, ,
\end{aligned}
\end{equation}
where $e(\mathbf{k}) = \sqrt{m_N^2 + \mathbf{k}^2}$.
The factors $m_{N}/e(\textbf{k})$ and $m_{N}/e(\textbf{k} + \textbf{q})$ are included to account for the covariant normalization of the four spinors in the matrix elements of the relativistic current. The energy transfer has been replaced by $\tilde{\omega} = \omega - m_{N} + E - e(\textbf{k})$ to account for scattering off of a bound nucleon. Finally, the calculation of the one-nucleon spectral function $P_{h}(\textbf{k},E)$ provides the probability of removing a nucleon with momentum $\textbf{k}$ and leaving the residual nucleus with an excitation energy $E$; its derivation will be discussed in Sec.~\ref{sec:SpectralFunc}.
\subsection{Two Body Currents}
\begin{figure}[h]
\centering
    \includegraphics[width=9cm]{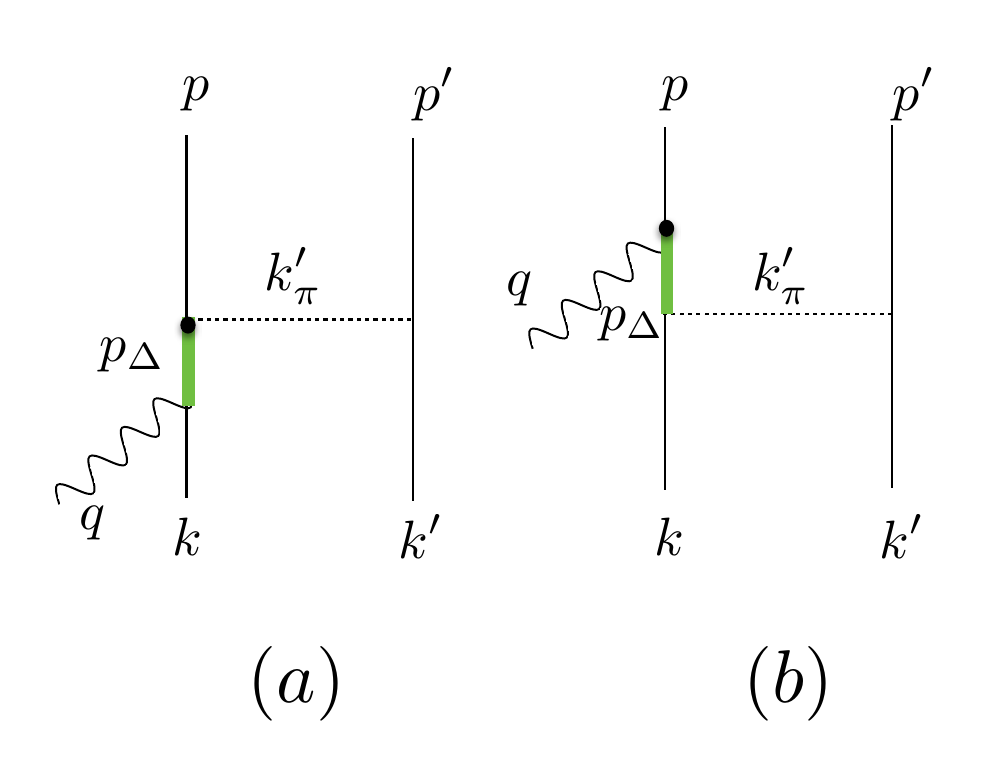} 
    \caption{Feynman diagrams describing the first two contributions to the two-body currents associated with $\Delta$-excitation processes. Solid, thick green, and dashed lines correspond to nucleons, deltas, pions, respectively. The wavy line represents the vector boson.}
    \label{fig:j_Delta}
\end{figure}

To describe amplitudes including two nucleon currents, the factorization ansatz of Eq.~\eqref{eq:fact} can be generalized as
\begin{align}
    |\Psi_{f}\rangle \rightarrow |p p^\prime \rangle_a \otimes |\Psi_f^{A-2}\rangle\, .
\end{align}
where $|p\,p^\prime\rangle_a=|p\,p^\prime\rangle-|p^\prime\,p\rangle$ is the anti-symmetrized state of two-plane waves with momentum $p$ and $p^\prime$. Following the work presented in Refs.~\cite{Benhar:2015ula,Rocco:2015cil,Rocco:2018mwt}, the pure two-body current component of the response tensor can be written as
\begin{align}
\label{eq:W2body}
&R^{\mu\nu}_{\mathrm 2b}({\bf q},\omega)=\frac{V}{2} \int dE \frac{d^3k}{(2\pi)^3}  \frac{d^3k^\prime}{(2\pi)^3}\frac{d^3p}{(2\pi)^3}
\frac{m_N^4}{e({\bf k})e({\bf k^\prime})e({\bf p})e({\bf p^\prime})} \nonumber\\
 &\qquad \times  P_h({\bf k},{\bf k}^\prime,E) \sum_{ij}\, \langle k\, k^\prime | {j_{ij}^\mu}^\dagger |p\,p^\prime\rangle_a \langle p\,p^\prime |  j_{ij}^\nu | k\, k^\prime \rangle \delta(\omega-E+2m_N-e(\mathbf{p})-e(\mathbf{p}^\prime))\, .
\end{align}
In the above equation, the normalization volume for the nuclear wave functions $V=\rho / A$ with $\rho=3\pi^2 k_F^3/2$ depends on the Fermi momentum of the nucleus, which for $^{12}$C is taken to be $k_F=225$ MeV. In previous calculations of the above two-body hadron tensor the 
two-nucleon spectral function $P_h({\bf k},{\bf k}^\prime,E)$ has been approximated as a product of two one-nucleon spectral functions (see \ref{sec:SpectralFunc} for a more detailed discussion). This is correct in the limit of infinite nuclear matter limit where the two-nucleon momentum distribution can be split according to
\begin{equation}
n(k,k') = n(k)n(k') + \mathcal{O}(1/A)\, .
\end{equation}
Going beyond this approximation for medium mass nuclei involves the full calculation of the two-nucleon spectral function including all correlations and will be discussed further in Sec.~\ref{sec:SpectralFunc}. The two body current operator in Eq.~\eqref{eq:W2body} is given by a sum of four distinct contributions, namely the pion in flight, seagull, pion-pole, and $\Delta$ excitations
\begin{align}
j^\mu= (j^\mu_\mathrm{pif})+(j^\mu_\mathrm{sea}) + (j^\mu_\mathrm{pole}) + (j^\mu_{\Delta})\, 
\label{eq:jijCC}
\end{align}
and dubbed as Meson Exchange Currents (MEC). 
Detailed expressions for each term in Eq.~\eqref{eq:jijCC} can be found in Refs.~\cite{Simo:2016ikv,Rocco:2018mwt}. Below, we only report the two-body current terms involving a $\Delta$-resonance in the intermediate state, as we find them to be the dominant contribution. Because of the purely transverse nature of this current, the form of its vector component is not subject to current-conservation constraints and its expression is largely model dependent, as discussed in Sec.~\ref{sec:ham_curr}. The current operator can be written as ~\cite{Hernandez:2007qq,Simo:2016ikv}:
\begin{align}
(j^\mu_\Delta)_{\mathrm CC}&=\frac{3}{2}\frac{f_{\pi NN} f^\ast}{m^2_\pi} \Bigg\{ 
\Bigg[ \Big( -\frac{2}{3}\tau^{(2)}_{\pm}+\frac{(\tau^{(1)}\times \tau^{(2)})_\pm}{3}\Big) F_{\pi NN}(k^\prime_\pi) F_{\pi N \Delta} (k^\prime_\pi) (j^\mu_{a})_{(1)} \nonumber \\
&-\Big(\frac{2}{3}\tau^{(2)}_{\pm}+\frac{(\tau^{(1)}\times \tau^{(2)})_\pm}{3}\Big) F_{\pi NN}(k^\prime_\pi) F_{\pi N \Delta} (k^\prime_\pi) (j^\mu_{b})_{(1)}\Bigg]\Pi(k^\prime_\pi)_{(2)}+(1\leftrightarrow 2)\Bigg\}\, ,
\label{delta:curr}
\end{align}
where $k^\prime$ and $p^\prime$ are the initial and final momentum of the second nucleon, respectively, while $k^\prime_\pi= p^\prime- k^\prime$ is the momentum of the $\pi$ exchanged in the two depicted diagrams of Fig.~\ref{fig:j_Delta}, $f^\ast$=2.14, and 
\begin{align}
&\Pi(k_\pi)=\frac{\gamma_5 \slashed{k}_\pi}{k_\pi^2-m^2_\pi}\ ,\\
&F_{\pi N \Delta}(k_\pi)=\frac{\Lambda^2_{\pi N\Delta}}{\Lambda^2_{\pi N\Delta}-k_\pi^2}\ , \\
&F_{\pi NN}(k_\pi)= \frac{\Lambda_\pi^2-m_\pi^2}{\Lambda_\pi^2-k_\pi^2}\label{fpinn}\ ,
\end{align}
with $\Lambda_{\pi N\Delta}=1150$ MeV and $\Lambda_{\pi}=1300$ MeV. 
In Eq.~\eqref{delta:curr}, $j^\mu_a$ and $j^\mu_b$ denote the $N\rightarrow \Delta$ transition vertices of diagram (a) and (b) of Fig.~\ref{fig:j_Delta}, respectively. The expression of $j^\mu_a$ is given by 
\begin{align}
j^\mu_a&=(j^\mu_a)_V+(j^\mu_a)_A\ ,\nonumber\\
(j^\mu_a)_V&=(k_\pi^\prime)^\alpha G_{\alpha\beta}(p_\Delta)\Big[\frac{C_3^V}{m_N}\Big(g^{\beta\mu}\slashed{q}-q^\beta\gamma^\mu\Big)+\frac{C_4^V}{m_N^2}\Big(g^{\beta\mu}q\cdot p_\Delta-q^\beta p_\Delta^\mu\Big)\nonumber\\
&+\frac{C_5^V}{m_N^2}\Big(g^{\beta\mu}q\cdot k-q^\beta k^\mu + C_6^V g^{\beta\mu}\Big)
\Big]\gamma_5\ ,\nonumber\\
(j^\mu_a)_A&= (k_\pi^\prime)^\alpha G_{\alpha\beta}(p_\Delta)\Big[ \frac{C_3^A}{m_N}\Big(g^{\beta\mu}\slashed{q}-q^\beta\gamma^\mu\Big) +\frac{C_4^A}{m_N^2}\Big(g^{\beta\mu}q\cdot p_\Delta-q^\beta p_\Delta^\mu\Big)+ C_5^A g^{\beta\mu}+\frac{C_6^A}{m_N^2}q^\mu q^\alpha \Big],
\label{eq:delta:curr:ja}
\end{align}
 where $k$ is the momentum of the initial nucleon which absorbs the incoming momentum $\tilde{q}$ and $p_\Delta =\tilde{q}+k$, yielding $p^0_\Delta=e({\bf k})+\tilde{\omega}$. We introduced $\tilde{q}=(\tilde{\omega},{\bf q})$ to account for the fact that the initial nucleons are off-shell. A similar definition can be written down for $j^\mu_b$; more details are reported in Ref.~\cite{Rocco:2019gfb,Rocco:2018mwt}. 
For $C_3^V$ we adopted the model of Ref.~\cite{Lalakulich:2006sw}, yielding
\begin{align}
    C_3^V= \frac{2.13}{(1-q^2/M_V^2)^2}\frac{1}{1-q^2/(4 M_V^2)},
\end{align}
with $M_V=0.84$ GeV. Following the discussion of Ref.~\cite{Simo:2016ikv}, we neglected the terms $C_4^V$ and $C_5^V$ which are expected to be suppressed by $\mathcal{O}(k/m_N)$, while $C_6^V=0$ by conservation of the vector current. However, it is worth mentioning that including these terms in the current operator would not pose any conceptual difficulty. 
 To be consistent, in the axial part we only retain the leading contribution of Eq.~\eqref{eq:delta:curr:ja}, which is the term proportional to $C_5^A$ defined as ~\cite{Paschos:2003qr}
 \begin{align}
     C_5^A= \frac{1.2}{(1-q^2/M_{A\Delta})^2} \times \frac{1}{(1-q^2/(3M_{A\Delta}))^2},
 \end{align}
 with $M_{A\Delta}= 1.05$ GeV. 

The Rarita-Schwinger propagator 
\begin{align}
G^{\alpha\beta}(p_\Delta)=\frac{P^{\alpha\beta}(p_\Delta)}{p^2_\Delta-M_\Delta^2},
\label{eq:free_delta}
\end{align}
is proportional to the spin 3/2 projection operator $P^{\alpha\beta}(p_\Delta)$.
In order to account for the possible decay of the $\Delta$ into a physical $\pi N$, we replace $M_\Delta \rightarrow M_\Delta - i \Gamma(p_\Delta)/2$~\cite{Dekker:1994yc,DePace:2003spn} where the last term is the energy dependent decay width given by
\begin{align}
\Gamma(p_\Delta)&=\frac{(4 f_{\pi N \Delta})^2}{12\pi m_\pi^2} \frac{|\mathbf{d}|^3}{\sqrt{s}} (m_N + E_d) R(\mathbf{r}^2)\, .
\label{eq:decay_width}
\end{align}
In the above equation, $(4 f_{\pi N \Delta})^2/(4\pi)=0.38$, $s=p_\Delta^2$ is the invariant mass, $\mathbf{d}$ is the decay three-momentum
in the $\pi N$ center of mass frame, such that
\begin{equation}
|\mathbf{d}|^2=\frac{1}{4s}[s-(m_N+m_\pi)^2][s-(m_N-m_\pi)^2]\, ,
\end{equation} 
and $E_d=\sqrt{m_N^2 + \mathbf{d}^2}$ is the associated energy. The additional factor
\begin{equation}
R(\mathbf{r}^2)=\left(\frac{\Lambda_R^2}{\Lambda_R^2-\mathbf{r}^2}\right) ,
\end{equation}
depending on the $\pi N$ three-momentum $\mathbf{r}$, with $\mathbf{r}^2=(E_d - \sqrt{m_\pi^2 + \mathbf{d}^2})^2-4\mathbf{d}^2$ and $\Lambda_R^2=0.95\, m_N^2$,
is introduced to improve the description of the experimental phase-shift $\delta_{33}$~\cite{Dekker:1994yc}.
The medium effects on the $\Delta$ propagator are accounted for by modifying the decay width as
\begin{align}
    \Gamma_\Delta(p_\Delta)\to \Gamma_\Delta(p_\Delta)-2\mathrm{Im}[U_\Delta(p_\Delta,\rho=\rho_0)],
\end{align}
where $U_\Delta$ is a density dependent potential obtained from a Bruckner-Hartree-Fock calculation using 
a coupled-channel $NN\oplus N\Delta\oplus \pi NN$ model~\cite{Lee:1983xu,Lee:1984us,Lee:1985jq,Lee:1987hd} and we fixed the density at the nuclear saturation value $\rho_0$ =0.16 fm$^3$. For a detailed analysis of medium effects in the MEC contribution for electron-nucleus scattering see Ref.~\cite{Rocco:2019gfb}.
One key point to be made from the point of inclusive and even semi-exclusive observables is that the one and two body currents contribute coherently, i.e. that their interference terms are non zero. The interference between one and two body currents leading to two-nucleon emission has been found to be small~\cite{Benhar:2015ula}, but the same interference also contributes to single nucleon final states~\cite{Fabrocini:1997,Franco-Munoz:2022jcl}. The impact of the latter interference in the SF formalism remains to be studied. 

\section{Spectral Function}\label{sec:SpectralFunc}
In the factorization scheme of Sec.~\ref{sec:fact}, the spectral function is the central object containing all the dynamical information about the nucleus. The spectral function of a nucleon with isospin $\tau_k=p,n$ and momentum ${\bf k}$ can be written as
\begin{align}
    P_{\tau_k}(\mathbf{k},E)&=\sum_n |\langle \Psi_{0}| [|k\rangle\otimes |\Psi_n^{A-1}\rangle]|^2 \delta(E+E_0-E_{n}^{A-1})\, .
\label{pke:hole}
\end{align}
where $E$ is the excitation energy of the remnant, $|k\rangle$ is the single-nucleon state, $|\Psi_{0}\rangle$ is the ground state of the Hamiltonian in Eq.~\eqref{eq:hamiltonian} with energy $E_0$, while $|\Psi_n^{A-1}\rangle$ and $E_{n}^{A-1}$ are the energy eigenstates and eigenvalues of the remnant nucleus with $(A-1)$ particles. The momentum distribution of the initial nucleon is obtained by integrating the spectral function over $E$ 
\begin{equation}\label{eq:momentum_dist}
n_{\tau_k}(k) = \int dE P_{\tau_k}(\mathbf{k},E)\, ,
\end{equation}
and the proton and neutron spectral functions are normalized so that
\begin{align}
        &\int dE\frac{d^{3}k}{(2\pi)^{3}}P_{p}(\mathbf{k},E) = Z\, ,\nonumber\\
        &\int dE\frac{d^{3}k}{(2\pi)^{3}}P_{n}(\mathbf{k},E) = A-Z\, .
\end{align}
We can rewrite the spectral function as a sum of a mean field (MF) and a correlation (corr) term. The MF piece contains the shell structure with nucleons occupying orbitals obeying the Pauli principle and contributes to the low $k$ and $E$ region. On the other hand, the correlation term comes from pairs and triplets of interacting nucleons with low center of mass momentum but large relative momentum above $k_{f}$. A large body of experimental evidence from $(e,e'p)$ data has shown that the correlation piece leads to a depletion of the single nucleon strength in the MF region by approximately 20\% and is essentially nucleus independent~\cite{CLAS:2005ola,CLAS:2022odn,Weiss:2020bkp,Hen:2014nza,JeffersonLabHallA:2007lly,CLAS:2020rue}.

Many calculations of the spectral function for finite nuclei are available from a combination of fits to $(e,e'p)$ cross sections and theoretical calculations. The spectral function of Benhar et al. obtains the mean field piece from fits to exclusive electron scattering data, and computes the correlation piece from CBF theory for nuclear matter~\cite{Benhar:1994hw,Benhar:1989aw}. The local density approximation (LDA) is used to extrapolate the correlation piece to finite nuclei by convoluting the correlation component of the nuclear spectral function $P_{\mathrm{NM}}^{\mathrm{corr}}$ with density profile of the nucleus $\rho_{A}(\mathbf{R})$~\cite{VanNeck:1994ge}, it reads
\begin{equation}
    P^{\mathrm{corr}}_{\mathrm{CBF}}(\mathbf{k},E) = \int d^{3}R\,\rho_{A}(\mathbf{R})P_{\mathrm{NM}}^{\mathrm{corr}}(\mathbf{k},E;\rho_{A}(\mathbf{R}))\, .
\end{equation}

In addition to the CBF, the spectral function of nuclear matter and finite nuclei has been computed within the
Self Consistent Green's Function approach. The latter is a so-called $\textit{ab initio}$ method method that starts from a nuclear Hamiltonian such as Eq.\eqref{eq:hamiltonian} with NNLO chiral interactions\cite{Dickhoff:2004xx,Barbieri:2016uib}. The SCGF method involves an iterative calculation of the Green's function's imaginary component, which is directly related to the one-body spectral function. This technique can be extended to open shell nuclei and has a polynomial scaling with the number of particles, making it feasible for systems with up to $A = 100$~\cite{Barbieri:2019ual}.
Both CBF and SCGF spectral functions have been used to compute inclusive electron and neutrino scattering cross sections, and have been shown to provide good agreement with electron data when final state interactions are taken into account~\cite{Rocco:2020jlx}. Even though the two spectral functions are obtained from different nuclear interactions, these calculations show that the two many-body approaches produce similar results for inclusive cross sections. Exclusive predictions will most likely be necessary to distinguish the two models. 

In this work we focus on a novel Quantum Monte Carlo (QMC) calculation of the one- and two-body spectral function for $^{12}$C. We begin with the MF piece of the one body SF for the $\tau_{k} = p$ case. The MF contribution is obtained by considering only bound $A-1$ states of the remnant nucleus
\begin{align}
P_p^{\mathrm MF} (\mathbf{k},E)&=\sum_n|\langle \Psi_{0}| [|k\rangle \otimes |\Psi_n^{A-1}\rangle]|^2 \nonumber\\
&\times \delta\Big(E-B^{A}_0+B_n^{A-1}-\frac{k^2}{2m_{A-1}}\Big)\, ,
\end{align}
where $B^{A}_0$ and $B_n^{A-1}$ are the binding energies of the initial and the bound $A-1$ spectator nucleus with mass $m_{A-1}$. The momentum-space overlaps $\Psi_{0} | [|k\rangle \otimes |\Psi_n^{A-1}\rangle$ pertaining to the p-shell contributions are computed by Fourier transforming the variational Monte Carlo (VMC) radial overlaps for the transitions ~\cite{Mecca:2019lxd,overlaps_web}: 
\begin{align*}
    ^{12}{\mathrm C}(0^+)&\rightarrow ^{11}{\mathrm B}(3/2^-)+p \nonumber\\
    ^{12}{\mathrm C}(0^+)&\rightarrow ^{11}{\mathrm B}(1/2^-)+p \nonumber\\
    ^{12}{\mathrm C}(0^+)&\rightarrow ^{11}{\mathrm B}(3/2^-)^\ast+p\, .
\end{align*}
The calculation of the s-shell mean-field contribution involves non trivial difficulties for the VMC method, as it would require to evaluate  the spectroscopic overlaps for the transitions to all the possible excited states of $^{11}{\mathrm B}$ with $J^P=(1/2^+)$. To overcome this limitation, we used the VMC overlap associated with the $^{4}\mathrm {He}(0^+)\rightarrow {}^{3}{\mathrm H}(1/2^+)+p$ transition and applied minimal changes to the quenching factor which is needed to reproduce the integral of the momentum distribution up to $k_F= 1.15$ fm$^{-1}$. More details about the adopted procedure are discussed in Ref.~\cite{CLAS:2022wvz}.  

The correlation contribution to the SF is given by
\begin{align}
P_p^{\mathrm corr}&(\mathbf{k},E)=\sum_n \int \frac{d^3 k^\prime}{(2\pi)^3} |\langle\Psi_{0}| [|k\rangle\,  |k^\prime\rangle \, |\Psi_n^{A-2}\rangle]|^2\nonumber\\
&\times \delta(E+E_0-e(\mathbf{k}^\prime)-E_{n}^{A-2})\, \nonumber\\
&=  \mathcal{N}_p \sum_{\tau_{k^\prime}=p,n} \int \frac{d^3 k^\prime}{(2\pi)^3} \Big[ n_{p,\tau_{k^\prime}}(\mathbf{k},\mathbf{k}^\prime)\nonumber \\
&\qquad \times \delta \Big(E-B_0-e(\mathbf{k}^\prime)+\bar{B}_{A-2}-
\frac{(\mathbf{k}+\mathbf{k}^\prime)^2}{2m_{A-2}}\Big)\Big]\, , 
\label{pke:2b}
\end{align}
to derive the last expression, we used a completeness relation and assumed that the $(A -2)$-nucleon binding energy is narrowly distributed around a central value $\bar{B}_{A-2}$. The mass of the recoiling $A-2$ system is denoted by $m_{A-2}$ and $\mathcal{N}_p$ is an appropriate normalization factor. 
We started from the VMC two-nucleon momentum distribution $n_{\tau_k,\tau_{k^\prime}}(\mathbf{k},\mathbf{k}^\prime)$ of Ref.~\cite{nkk_web}, but in order to isolate the contribution of short-range correlated nucleons we performed cuts in the relative momentum of the pairs, requiring that the overall normalization and shape of the one-nucleon momentum distributions are correctly recovered. Below in Fig.~\ref{fig:single_spectral_func} we show the $^{12}$C single nucleon momentum distribution derived using the above procedure. 
\begin{figure}[h]
\includegraphics[width=13cm]{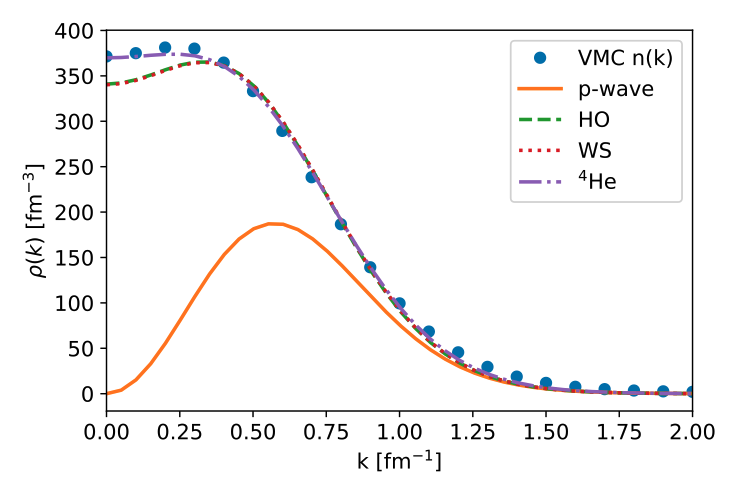} 
\caption{$^{12}$C Single proton QMC spectral function (blue points). The solid orange line shows the sum of the p-wave overlaps between the $^{12}C$ and $^{11}$B+p QMC wave functions. The momentum distributions obtained by adding to the p-wave overlaps the different prescription for the s-wave contribution are displayed by the green dashed line (harmonic oscillator), dotted red line (Wood-Saxon) and dash-dotted purple line (s-wave overlaps between $^{4}$He and the $^{3}$H+p QMC wave functions). The high momentum contributions of long- and short-range correlations are not visible on this linear scale}
\label{fig:single_spectral_func}
\end{figure}
Figure~\ref{fig:single_spectral_func} shows the effect of the different prescriptions for calculating the s-wave overlaps, with the above prescription resulting in an  increased normalization of the SF compared to the harmonic oscillator and Wood-Saxon potentials. In Fig.~\ref{fig:C12_mom} we directly compare the QMC and CBF $^{12}$C spectral functions by comparing their one dimensional momentum and removal energy distributions. While the two SF have very similar removal energy distributions, their momentum distributions show distinct behavior at small and large nucleon momenta. Although these discrepancies only cause minor variations in the inclusive cross section, it is anticipated that they will be more significant in exclusive cross sections where the outgoing nucleon is measured. This will be explored in future studies.
\begin{figure}[h]
\includegraphics[width=14cm]{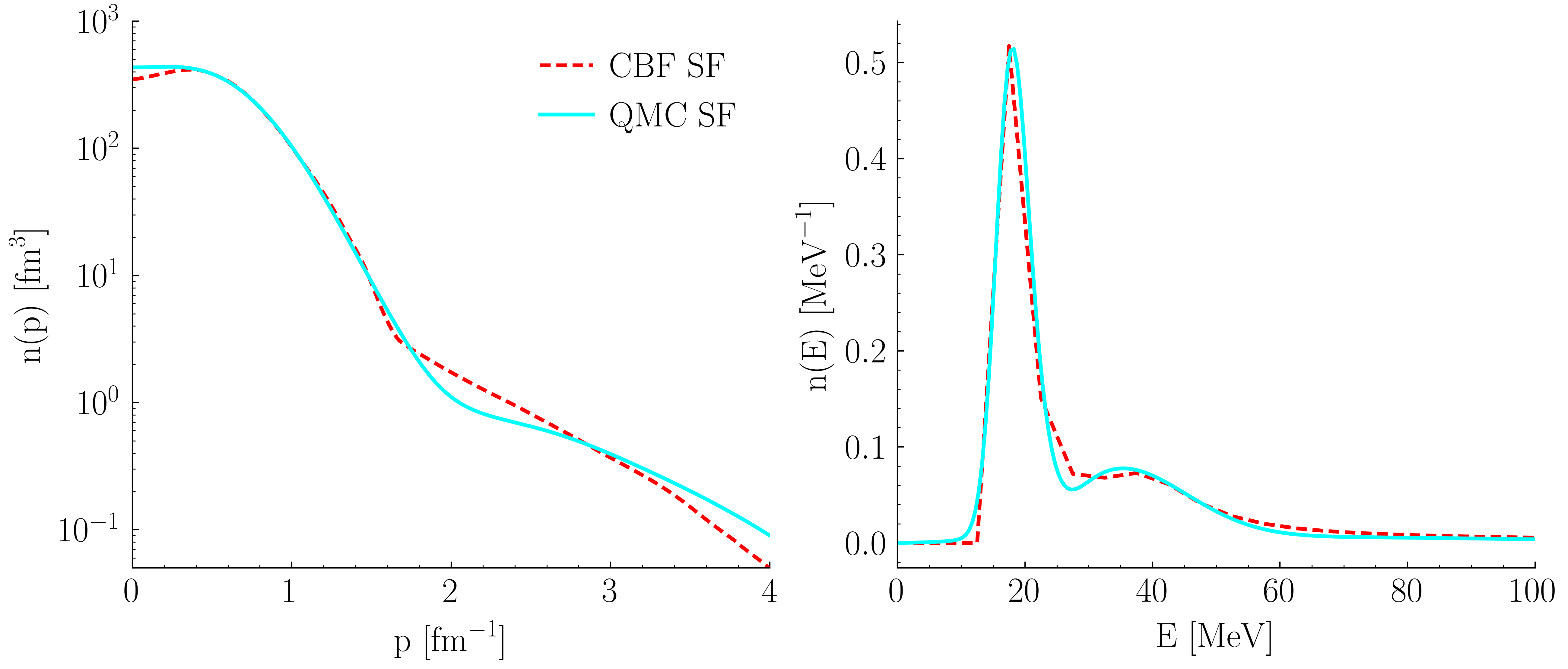} 
\caption{Comparison of nucleon momentum distribution (left) and removal energy distribution (right) in $^{12}$C using QMC techniques (solid cyan) vs. CBF theory (dashed red).}
\label{fig:C12_mom}
\end{figure}

For the contribution of multi-nucleon currents to the cross section, a two nucleon spectral function is needed. As mentioned previously, in infinite nuclear matter a factorization of the two nucleon momentum distribution into the product of two single nucleon momentum distributions can be made. This factorization assumes no long-range correlations are present and throws away correlations between the two struck particles. We go beyond this approximation by explicitly by using the two-nucleon spectral momentum distribution to build the two-nucleon spectral function. We include only the mean field contribution, e.g. we neglect contributions where more than two nucleons are emitted which reads 
\begin{align}
P_{\tau_k,\tau_k^\prime}^{\mathrm MF}(\mathbf{k},\mathbf{k}^\prime,E)&=
n_{\tau_k,\tau_{k^\prime}}(\mathbf{k},\mathbf{k}^\prime)\nonumber \\
& \times \delta \Big(E-B_0+\bar{B}_{A-2}-
\frac{\mathbf{K}^2}{2m_{A-2}}\Big)\, , 
\end{align}
where $\mathbf{K}=\mathbf{k}+\mathbf{k}^\prime$ is the total momentum of the pair.

\section{GFMC and SF Comparisons}
Recently the authors of Ref.~\cite{Simons:2022ltq} computed flux folded differential cross sections for MiniBooNE and T2K experiments ~\cite{Aguilar-Arevalo:2004wet,t2k_web} using the QMC spectral function outlined above. Under control systematics for the calculation of the two-body contribution is required for disentangling the effect of the axial form factor. The QMC spectral function can be directly compared with GFMC predictions because it is derived from the same underlying Hamiltonian, and currents. Comparisons with experiments have shown that the predictions are consistent with the data and show tension between the results obtained adopting the LQCD and phenomenological form factors displayed in Fig.~\ref{fig:form_factors}~\cite{Meyer:2022mix}.
\begin{figure}[h]
\includegraphics[width=13cm]{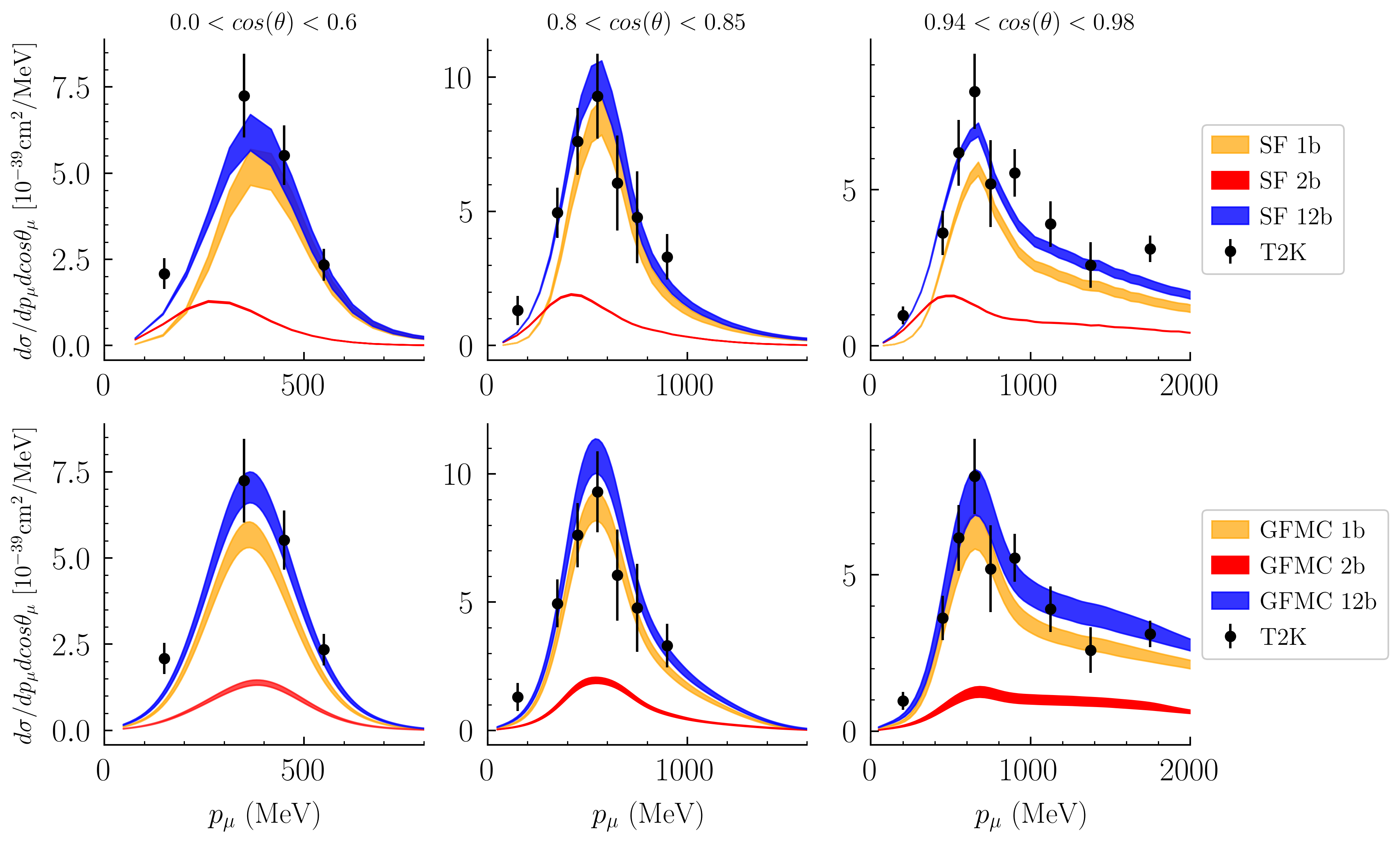} 
\caption{ Comparison with T2K data, adapted from Ref.~\cite{Simons:2022ltq}.
Breakdown into one- and two-body current contributions of the $\nu_{\mu}$ flux-averaged differential cross sections for T2K: 1b and 2b denotes one- and two-body current contributions while 12b denotes the total sum of these contributions. The top panel shows QMC SF predictions in three bins of $\cos\theta_{\mu}$ with the one-body contributions in orange, two-body contributions in red, and the total in blue. The lower panel shows GFMC predictions with the same breakdown between one- and two-body current contributions, although the two-body results include interference effects only in the GFMC case.  The D2 Meyer et al. $z$ expansion results for $F_{A}$ are used in both cases~\cite{Meyer:2016oeg}.}
\label{fig:T2K}
\end{figure}
Results for selected angular bins for T2K kinematics are shown in Fig.~\ref{fig:T2K}. The shown uncertainty bands propagate the uncertainty on the axial nucleon form-factor derived using the $z$-expansion where the different coefficients have been fitted to deuterium bubble chamber data of Ref.~\cite{Meyer:2016oeg}.
In the case of the GFMC, the uncertainty coming from the inversion of Euclidean responses is also included. 
Both approaches provide a similar description of the data, albeit the contribution of two-body currents peaks is shifted in the two approaches. This can be ascribed to different motivations. 
Firstly, the SF results include explicitly the contribution of $\Delta$-excitations in the  two-nucleon knockout process, leading to a peak at smaller lepton momenta, while the GFMC results use a static $\Delta$ treatment. 
Secondly, the GFMC results also account for the interference between the two-body and one-body currents, which would lead to an enhancement also in the vicinity of the quasi-elastic peak.
Such an enhancement is clearly seen in Fig.~\ref{fig:T2K}, and in the electromagnetic and electroweak responses~\cite{PhysRevX.10.031068, PhysRevLett.117.082501}. 
While these observations support the one- and two-body current interference, it is impossible to disentangle this contribution directly in the GFMC results. The calculations in nuclear matter and relativistic mean field calculations of Refs.~\cite{Fabrocini:1997,Franco-Munoz:2022jcl} also find that the transverse enhancement observed in electron scattering is primarily due to the constructive interference between one- and two-body currents, leading to single-nucleon knockout final states.   

Recently~\cite{Nikolakopoulos:2023zse}, relativistic corrections to GFMC calculations for flux-averaged neutrino cross sections has been determined using the method described in Sec.~\ref{sec:relcor}. The influence on T2K results shown in Fig.~\ref{fig:T2K}, is small and generally falls within the uncertainty bands due to the axial form factor. 
For MINER$\nu$A data~\cite{MINERvA:2023kuz} taken with the medium-energy NuMI beam, which peaks at around $6~\mathrm{GeV}$~\cite{NUMI:ME2019}, relativistic corrections are crucial.
The charged-current flux-averaged cross section is presented in terms of muon momentum parallel and perpendicular to the beam direction
\begin{equation}
    p_{\parallel} =  \lvert \mathbf{p}_\mu \rvert \cos\theta_\mu ,
\end{equation}
and
\begin{equation}
    p_{\perp} = \lvert \mathbf{p}_\mu \rvert \sin\theta_\mu = \sqrt{\mathbf{p}_\mu ^2 - p_\parallel^2 }, 
\end{equation}
respectively.

\begin{figure}[h]
    \includegraphics[width=14cm]{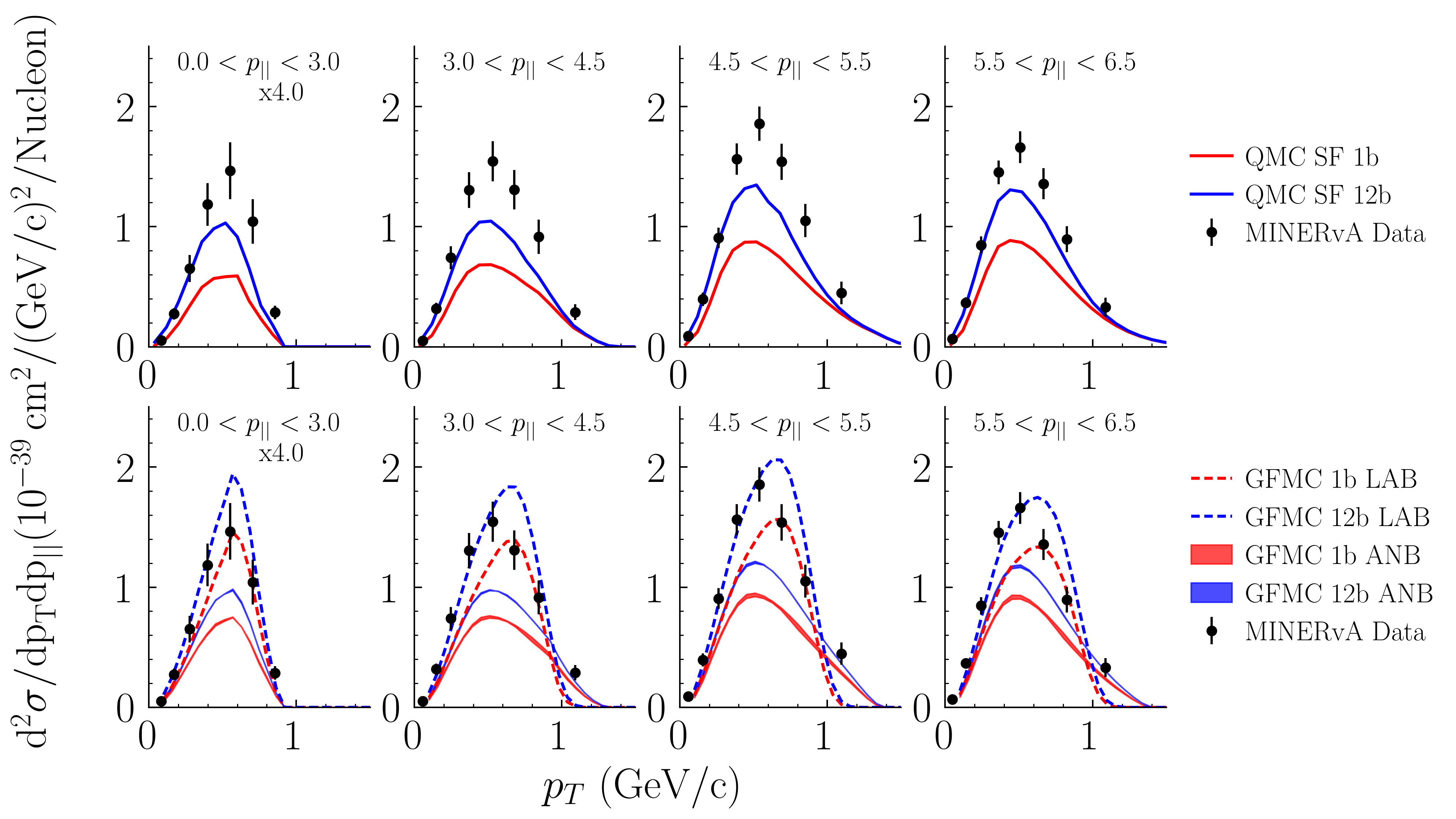} 
    \caption{Comparison with MINERvA Medium Energy CCQE-like data on CH. Cross section per nucleon is measured double differential against $p_{T}$ (momentum transverse to beam direction) in bins of $p_{||}$ (momentum parallel to beam direction). Top panels show QMC SF prediction broken down into one-body (red) and one+two-body (blue) in different bins of $p_{||}$. Bottom panels show GFMC predictions again broken down into one- and one+two-body results,  with response functions computed in the LAB frame (dashed lines) and ANB frame (solid line). This thickness of the ANB curves corresponds to the error from the inversion procedure.}
    \label{fig:MINERvA}
\end{figure}

{ In this comparison, the routinely used dipole parameterization of the axial form factor with $\Lambda_A=1.03$ GeV has been used. For the GFMC results, the error band includes statistical errors combined with the error from the inversion of Euclidean responses. The SF results do not include an error estimate.}
Relativistic corrections are included in the GFMC results by performing the calculation in the active-nucleon Breit frame (ANB) as discussed in Sec.~\ref{sec:relcor}. {
The incorporation of relativistic effects leads to a nearly halved cross section for low-$p_\parallel$, with the discrepancy gradually decreasing as $p_\parallel$ increases. We observe that the momentum transfer is constrained such that $q > p_{\perp}$, and smaller $p_\parallel$ bins generally permit higher energy and consequently larger $q$ contributions at small $p_{\perp}$, thus explaining this behavior. The emergence of high-$p_\perp$ (i.e., high-$q$) tails can be understood as the response's narrowing in terms of energy transfer compared to nonrelativistic outcomes, resulting in strength redistribution within the available phase space at large-$q$. \\
Given the inclusion of large $q$ values in the MINER$\nu$A calculations and the substantial impact of relativistic corrections, a consistency check is warranted. In Ref. ~\cite{Nikolakopoulos:2023zse}, the GFMC results for MINER$\nu$A kinematics obtained including only the one-body current contribution have been compared to other approaches that are either manifestly relativistic~\cite{SuSav2} or include relativistic corrections~\cite{Pandey:2016, PRC92, Jachowicz:EPST, Dolan:2022CRPA} and found to agree with the theoretical curves. Here, Fig.~\ref{fig:MINERvA} compares the GFMC results to the SF calculations including both the one- and two-body contributions in Fig.~\ref{fig:MINERvA}.
The agreement between the one-body contribution in the GFMC and SF approaches is evident when the former are computed in the ANB. 
The total increase of the cross section due to two-body contributions is twice as large in the SF calculations compared to the GFMC. This difference can be attributed to the same motivations discussed for the T2K results. 

Lastly, we note that the GFMC nonrelativistic calculations exhibit better conformity with experimental data compared to those incorporating relativistic effects. However, considering the energy distribution of the medium-energy NuMI beam in the MINER$\nu$A experiment, it is expected that contributions beyond quasi-elastic scattering are significant, even when events with detectable mesons are excluded from the experimental analysis. Specifically, there are instances where pions produced at the interaction vertex are either absorbed or remain undetected. Thus, theoretical calculations that neglect pion-production mechanisms should yield results lower than experimental data. This aligns with the case where relativistic effects are considered, while their omission leads to un-physically large cross sections.}

\begin{figure}[h]
    \includegraphics[width=14cm]{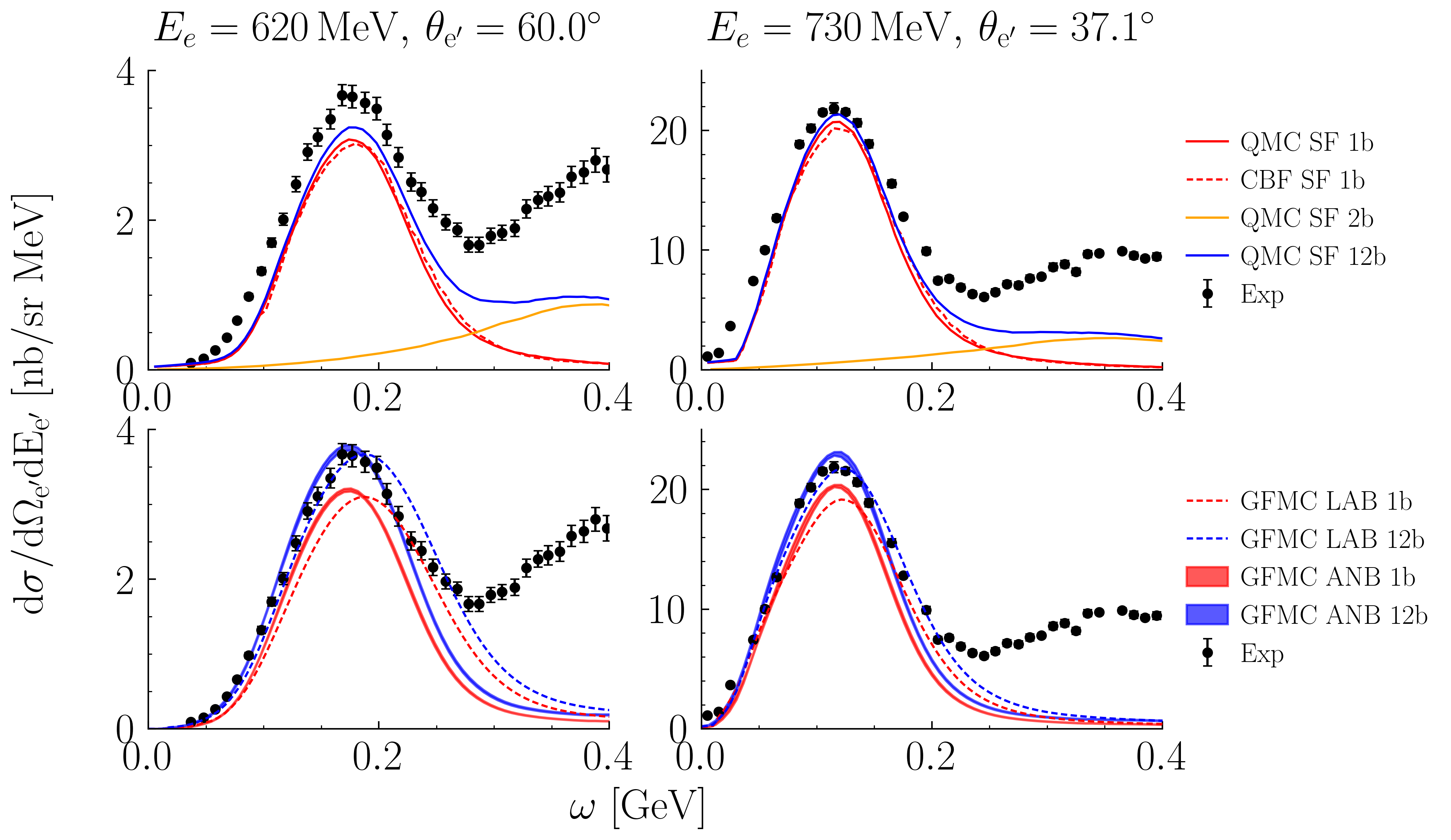} 
    \caption{Inclusive electron scattering comparisons at two different kinematics. Left: $E_{\mathrm{beam}} = 620\,\mathrm{MeV}$, $\theta_{e'} = 60^{\circ}$. Right: $E_{\mathrm{beam}} = 730\,\mathrm{MeV}$, $\theta_{e'} = 37.1^{\circ}$. Data is from Refs.~\cite{Sealock:1989nx, Barreau:1983ht, Benhar:2006er}. Upper panels are for SF with QMC (CBF) one body in solid (dashed) red, QMC two-body in orange, and QMC one+two-body in blue. GFMC predictions are in the lower panel with dashed lines corresponding to response functions computed in the LAB frame, and solid for response functions in the ANB frame. Error bars on GFMC calculations include only errors from the inversion of the Euclidean response function, but neglect uncertainty due to interpolation of the responses as discussed in the text.}
    \label{fig:electron_scattering}
\end{figure}

In this review, we also consider inclusive electron scattering data on $^{12}$C in Fig.~\ref{fig:electron_scattering} which allows one to disentangle the different energy regions more clearly. { The two kinematics under consideration have been deliberately selected to include only responses with $q < 700$ MeV. These specific values align with the range for which the GFMC responses have already been computed. Similarly to the neutrino case, in the GFMC calculations two-body currents provide an enhancement in the quasi-elastic region. 
A comparison between LAB frame results using purely relativistic kinematics (depicted by the blue dotted curve) and the ANB curve (solid blue) reveals important insights. Relativistic corrections cause a shift of the peak towards smaller $\omega$ values, a reduction in width, and an increase in the height of the quasi-elastic peak. The one-body contribution computed in the ANB frame, displayed by solid red line, agrees fairly well with the SF one-body contribution displayed in the upper panels. Overall agreement with the data improves by including relativistic corrections to the GFMC results. However, it is worth noting that the absence of $\pi$-production contributions makes it difficult to draw definitive conclusions without considering that term.

The static treatment of the $\Delta$ propagator restricts the significance of two-body currents in the "dip" region, located between the quasi-elastic and pion-production peaks. Incorporating explicit dynamical degrees of freedom in GFMC calculations is more challenging, particularly in terms of evaluating the Euclidean responses while fixing the current operator's dependence at multiple values of $\omega$.

The total QMC SF results encompass the incoherent sum of one-nucleon and two-nucleon contributions. We include in the calculation the effect of final state interactions by convoluting the computed cross sections with a folding function which both shifts and redistributes strength from the peak to the tails~\cite{Benhar:2006wy}. Two-body currents give a minor enhancement in the quasi-elastic peak region, but a strong enhancement in the "dip" region. Additionally, we include CBF SF results for the one-body cross section for comparison, which show similar trends to the QMC one-body cross section as expected. However, the QMC SF result notably under-predicts the data in the region of the quasi-elastic peak at $E_{\mathrm{beam}} = 620,\mathrm{MeV}$. Investigating the interference between one and two-body currents and its impact on these results will be a subject of future investigation. \\ As a general remark, one can choose to apply either the GFMC or the spectral function approach depending on the kinematics and process under investigation. The selection depends on the specific requirements of the study. However, it is important to ensure that the results obtained from both methods are consistent in the transition regions where both approaches are expected to work. }

\section{Conclusion}
Neutrino oscillation experiments cover a broad range of energies, from a few MeV to tens of GeV, where different reaction mechanisms involving various degrees of freedom (nucleons, pions, quarks, etc.) are active. Microscopic approaches such as Green's Function Monte Carlo (GFMC) and Coupled Cluster have been successful in describing lepton-nucleus cross-sections in the MeV energy region~\cite{Rocco:2017,Sobczyk:2020qtw,Sobczyk:2021dwm}. However, to address the higher energies relevant for DUNE and include explicit pion degrees of freedom, different methods relying on a factorization of the hadronic final state, such as the Spectral Function (SF), the Short Time Approximation~\cite{Pastore:2019urn,Andreoli:2021cxo}, and the Relativistic Mean Field approach~\cite{Gonzalez-Jimenez:2019qhq,SuSav2}, have proven successful in reproducing electron scattering data for different kinematics.

Providing a realistic estimate of the theoretical uncertainty of the prediction in the neutrino-nucleus cross-section, which must be propagated in the extraction of neutrino oscillation parameters, requires assessing the error associated with the input used in the calculations and with the many-body method used. In this review, we highlight that different choices can be made to define the nuclear forces adopted to describe the wave function of the target and remnant nucleus, either using semi-phenomenological approaches or chiral effective field theories. Following the choice of the nuclear forces, different current operators can also be constructed. Another source of uncertainty is connected to the form factors entering these currents. 
{ In Ref.~\cite{Simons:2022ltq} a study of the dependence of the neutrino-nucleus cross section results from the axial form factor adopted in the one-body current operator has been carried out using the GFMC and the SF approaches and a tension between the results obtained the LQCD and phenomenological form factors has been observed. The results of Ref.~\cite{Simons:2022ltq} indicate that, while significant progress has been made in the determination of the axial and vector form factors entering in the one-body current operator, more work will be required in the future for the determination of the form factors entering the two-body currents, particularly for those contributions with $\Delta$-degrees of freedom.  }

A two-fold strategy can be employed to comprehend the error associated with using a factorization scheme in the spectral function approach and nonrelativistic kinematics in GFMC. Firstly, relativistic corrections can be incorporated by working in a reference frame that minimizes them in the GFMC responses~\cite{Rocco:2017,Nikolakopoulos:2023zse}. Secondly, Quantum Monte Carlo (QMC) techniques can be used to derive one- and two-nucleon spectral functions. Comparing the results obtained from these two approaches can help estimate the error associated with the many-body method. Numerous studies have investigated this comparison. In this review, we present unpublished results that demonstrate electron-carbon cross-section comparisons and neutrino-nucleus cross-sections for the MINER$\nu$A experiment.
In the comparison with MINER$\nu$A Medium Energy CCQE-like data, the effect of relativistic corrections to the GFMC results are substantial, yielding a quenching of the results up to 50\% of the initial strength. We observe a reasonable agreement between the GFMC and QMC SF results. For the electron scattering cross section, we also analyzed the dependence of the results from the many-body method adopted to derive the spectral function, in particular, we compared the QMC and Correlated Basis Function results and found a very good agreement between them. Looking at fixed energy beam allows one to better separate the contribution of the different reaction mechanisms. In this case, the difference between the two-body contributions obtained within the two approaches is apparent and it has to be attributed to the different treatment of the $\Delta$-propagator in the GFMC and the lack of one- and two-body current interference in the SF approach. The inclusion of relativistic corrections in the GFMC results leads to better agreement with data. As there is a large amount of electron scattering data in the region of $300 < q < 700$ MeV, future studies that directly compare the GFMC results with differential electron scattering data for carbon can be performed.
A robust method for estimation of the uncertainty able to account for all the different aspect of the calculation is required to match the unprecedented accuracy of neutrino experiments, some preliminary steps toward this direction have been discussed in this review using the GFMC and SF methods.

\acknowledgments{
This manuscript has been authored by Fermi Research Alliance, LLC under Contract No. DE-AC02-07CH11359 with the U.S. Department of Energy, Office of Science, Office of High Energy Physics, Fermilab LDRD awards (N.S) and by the NeuCol SciDAC-5 program (N.R. and A.L.). The present research is also supported by the U.S. Department of Energy, Office of Science, Office of Nuclear Physics, under contracts DE-AC02-06CH11357 (A.L.), by the NUCLEI SciDAC-5 program (A.L.), the DOE Early Career Research Program (A.L.), and Argonne LDRD awards (A.L.). Quantum Monte Carlo calculations were performed on the parallel computers of the Laboratory Computing Resource Center, Argonne National Laboratory, the computers of the Argonne Leadership Computing Facility via INCITE and ALCC grants.}

\bibliography{reviewbib}

\end{document}